\documentclass[twocolumn]{svjour3}          % twocolumn
\smartqed  % flush right qed marks, e.g. at end of proof
\usepackage{graphicx}
\usepackage{mathptmx}      % use Times fonts if available on your TeX system
\usepackage{cite}
\usepackage{amsmath}
\usepackage{amssymb}
\usepackage{mathtools}
\usepackage{microtype}
\usepackage[caption=false]{subfig}
\usepackage{epstopdf}
\usepackage[dvipsnames]{xcolor}
\newcommand{\pts}{$\cal{PT}$}
\newcommand{\cps}{$\cal{CP}$}
\usepackage[breaklinks=true,colorlinks=true,linkcolor=blue,urlcolor=blue,citecolor=blue]{hyperref}
\journalname{Nonlinear Dynamics}

\begin{document}
\title{Discrete solitons dynamics in \pts-symmetric oligomers with complex-valued couplings}
\titlerunning{Discrete soliton dynamics in \pts-symmetric oligomers}
\author{O. B. Kirikchi \and  N. Karjanto}
\institute{O. B. Kirikchi \at
Department of Computing, Goldsmiths, University of London, New Cross, London SE14 6AD, United Kingdom \\
\and
N. Karjanto \at
Department of Mathematics, University College, Natural Science Campus, Sungkyunkwan University, 2066 Seobu-ro, Jangan-gu, Suwon 16419, Gyeonggi-do, Republic of Korea\\
\email{natanael@skku.edu}}

\date{Updated \today.}
\maketitle

\begin{abstract}
We consider an array of double oligomers in an optical waveguide device. A mathematical model for the system is the coupled discrete nonlinear Schr\"odinger (NLS) equations, where the gain-and-loss parameter contributes to the complex-valued linear coupling. The array caters to an optical simulation of the parity-time (\pts)-symmetry property between the coupled arms. The system admits fundamental bright discrete soliton solutions. We investigate their existence and spectral stability using perturbation theory analysis. These analytical findings are verified further numerically using the Newton-Raphson method and a standard eigenvalue-problem solver. Our study focuses on two natural discrete modes of the solitons: single- and double-excited-sites, also known as onsite and intersite modes, respectively. Each of these modes acquires three distinct configurations between the dimer arms, i.e., symmetric, asymmetric, and antisymmetric. Although both intersite and onsite discrete solitons are generally unstable, the latter can be stable, depending on the combined values of the propagation constant, horizontal linear coupling coefficient, and gain-loss parameter.

% Kerr/Photorefractive effect in nonlinear optics, nonlinear optics, self-focusing/self-phase modulation/beam trapping in nonlinear optics, nonlinear dynamics of solitons, classical mechancics of discrete systems, parity symmetry (fields and particles), perturbation theory applied to classical mechanics,

% Nonlinear waves, solitary waves, waves and radiation in optics, nonlinear optics, asymptotic analysis in optics, NLS-like equations, asymptotic behavior of soliton solutions, stability problems, soliton solutions
\end{abstract}

\section{Introduction}

Dissipative media featuring the parity-time (\pts)-symmetry has drawn a great deal of attention ever since Carl Bender and his collaborators proposed the system during the late 1990s~\cite{bender1998real, bender1999pt, bender2002complex, bender2007making}. The condition for a system of nonlinear evolution equations to be \pts-symmetry is that it is invariant with respect to both parity ${\cal P}$ and time-reversal ${\cal T}$ transformations. This type of symmetry is fascinating since it forms a specific family of non-Hermitian Hamiltonians in quantum physics that will possess a real-valued spectrum until a fixed parameter value of its corresponding complex potential. Above this critical value, the system then belongs to the broken \pts-symmetry phase~\cite{bender2007making, moiseyev2011non, kottos10broken, scott11broken}.

We assume that observable quantities in quantum mechanics are the eigenvalues of operators representing the dynamics of those quantities. Consequently, the eigenvalues, which epitomize the energy spectra, should be real-valued and acquire a lower bound to guarantee that the system features a stable lowest-energy state. To appease this requirement, we contemplate that the operators must be Hermitian. Non-Hermitian Hamiltonians are generally associated with complex-valued eigenvalues and thus degenerate the quantities. Interestingly, it turns out that the Hermiticity is not necessarily required by a Hamiltonian system to satisfy the Postulates of Quantum Mechanics~\cite{moiseyev2011non}. A necessary condition for a Hamiltonian to be \pts-symmetric is that its potential $V(x)$ should satisfy the condition $V (x) = V^{*}(-x)$~\cite{pickton2013integrability}.

The term ``oligomer'' is more well-known in the field of chemistry and comes from the Greek prefix \emph{oligo-}, ``a few'' and suffix \emph{-mer}, ``parts''. In this paper, it refers to a repeating structure composed of electronic oscillators or optical waveguides. A dimer is an oligomer system of two coupled oscillators, and it forms the most basic configuration of a system with a \pts-symmetry property. J{\o}rgensen and colleagues are the first authors who studied dimer and discussed the conditions for its integrability acquiring a Hamiltonian structure~\cite{jorgensen1993modified,jorgensen1994hamiltonian}.

A distinctive feature of the \pts-symmetry system is one part of the dimer loses energy due to a damping effect while another oscillator gains energy from an external source. Indeed, the idea of \pts-symmetry was accomplished experimentally for the first time on dimers consisting of two coupled optical waveguides~\cite{guo2009observation, ruter2010observation}. Optical analogs using two coupled waveguides with gain and loss were investigated in~\cite{ruschhaupt2005physical, el2007theory, klaiman2008visualization}, where such couplers have been considered previously in the 1990s~\cite{chen1992twin, jorgensen1993modified, jorgensen1994hamiltonian}.

\pts-symmetric analogs in coupled oscillators have also been proposed theoretically and experimentally~\cite{schindler2011experimental, ramezani2012bypassing, lin2012experimental, schindler2012mathcal}. A \pts-symmetric system of coupled oscillators with gain and loss can form a Hamiltonian system and exhibits a twofold transition which depends on the size of the coupling parameter~\cite{bender2013twofold,bender2014systems,barashenkov2014exactly}. A comparison between analytical study and numerical approach in a \pts-system with periodically varying-in-time gain and loss modeled by two coupled Schr\"odinger equations shows a remarkable agreement~\cite{battelli2015dynamics}. Besides showing that the problem can be reduced to a perturbed pendulum-like equation, they also investigated an approximate threshold for the broken \pts-symmetry phase.

In the case of the anticontinuum limit, breathers are common occurrences in the \pts-symmetric chain of dimers. Particularly, a system of amplitude equations governing the breather envelope remains conservative and the small-ampli-tude \pts-breathers are stable for a finite time scale~\cite{barashenkov2012breathers}. There exists a fascinating class of optical systems where a coupling or interaction causes the systems to be \pts-symmetric. Additionally, symmetry-breaking bifurcations in specific reciprocal and nonreciprocal \pts-symmetric systems have a promising application in optical isolators and diodes~\cite{karthiga2016systems}.

In addition to the \pts-symmetry phase transition, the reciprocal transmission and unidirectional reflectionless features are appealing to many. The axial and reflection \pts-symmetry lead to symmetric reflection and symmetric transmission, respectively~\cite{jin2016reciprocal}. Two interesting nonreciprocal phenomena are unidirectional lightwave propagation and unidirectional lasing, where both are independent of the input direction. When they are combined in a \pts-symmetric setting, the unidirectional destructive interference plays an important role in wave dynamics due to the vanishing of spectral singularity~\cite{jin2018incident}.

In particular, we are interested in the nonlinear dynamics of \pts-symmetric chain of dimers that can be modeled by the discrete nonlinear Schr\"{o}dinger (DNLS) type of equations due to its abundance applications in nonlinear optics and Bose-Einstein condensates (BEC) \cite{kevrekidis2001discrete,kivshar2003optical,kevrekidis2009discrete}. Transport on dimers with \pts-symmetric potentials are modeled by the coupled DNLS equations with gain and loss, which was relevant among others to experiments in optical couplers and proposals on BEC in \pts-symmetric double-well potentials~\cite{chernyavsky2016}. This proposed model is integrable and its integrability is further utilized to build up the phase portrait of the system. The existence and stability of localized mode solutions to nonlinear dynamical lattices of the DNLS type of equations with two-component settings have been considered and a general framework has been provided in~\cite{li2012intrinsic}. A dual-core nonlinear waveguide with the \pts-symmetry has been expanded by including a periodic sinusoidal variation of the loss-gain coefficients and synchronous variation of the inter-core coupling constant~\cite{fan2019dynamical}. The system leads to multiple-collision interactions among stable solitons. A study of the nonlinear nonreciprocal dimer in an anti-Hermitian lattice with cubic nonlinearity has been explored recently~\cite{tombuloglu2020nonlinear}.

In our previous work, we have considered the existence and linear stability of fundamental bright discrete solitons in \pts-symmetric dimers with gain-loss terms~\cite{kirikchi2016bright}, in a chain of charge-parity $({\cal CP})$-symmetric dimers~\cite{kirikchi2018solitons}, and in a chain of \pts-symmetric dimers with cubic-quintic nonlinearity~\cite{susanto2018snakes}. The latter covers the snaking behavior in the bifurcation diagrams for the existence of standing localized solutions. In this paper, we consider the coupled discrete linear and nonlinear Schr\"{o}dinger equations on oligomers with complex couplings as systems of \pts-symmetric potentials. This proposed model arises as nonlinear optical waveguide couplers or a BEC emulation in double-well potentials with \pts-symmetry and we hope to stimulate a series of experiments along this direction.

This article is organized as follows. We introduce the corresponding equations modeling the dynamics of the \pts-symmetric chain of dimers in Section~\ref{mathmodel}. We investigate the existence of fundamental discrete solitons for a small value of the linear horizontal coupling between two adjacent sites using the theory of perturbation. We analyze the stability of the fundamental discrete solitons by solving the corresponding eigenvalue problems in Section~\ref{stability}. Since the expression of the corresponding eigenvectors from the linearized eigenvalue problem is arduous, we also employ perturbation expansion with respect to the gain-loss parameter, which also assumed to be small. Section~\ref{numerical} compares analytical findings with numerical calculations. We display the plots of the spectra and typical dynamics of discrete solitons for different parameter values. Section~\ref{conclusion} concludes our study.

% Section 2
\section{Mathematical model}  			\label{mathmodel}

The coupled discrete NLS equations that govern the dynamics of \pts-symmetric chains of dimers are given as follows:
\begin{equation}
\begin{aligned}		 		\label{1model}
\dot{u}_{n} &= i|u_{n}|^{2}u_{n} + i\epsilon \Delta_{2} u_{n} + \gamma v_{n} + iv_{n},	\\ 
\dot{v}_{n} &= i|v_{n}|^{2}v_{n} + i\epsilon \Delta_{2} v_{n} - \gamma u_{n} + iu_{n},
\end{aligned}
\end{equation}
where the dots represent the derivative with respect to the evolution variable, which is the physical time $t$ for BEC and the propagation direction $z$ in the case of nonlinear optics. Both $u_{n} = u_{n}(t)$ and $v_{n} = v_{n}(t)$ are complex-valued wave functions at the site $n \in \mathbb{Z}$. The coefficient $0 < \epsilon << 1$ acts as the linear horizontal coupling constant between two adjacent sites. The quantities following $\epsilon$ are the discrete Laplacian factors in one spatial dimension, explicitly given as $\Delta_{2}u_{n} = (u_{n+1} - 2 u_{n} + u_{n-1})$ and $\Delta_{2}v_{n} = (v_{n+1} - 2 v_{n} + v_{n-1})$. The coefficient $\gamma$ represents the gain and loss factor and contributes to the complex-valued coupling of the system. Without loss of generality, we will take $\gamma > 0$. We only consider the solitons solutions that satisfy the localization conditions, i.e., $u_{n}, v_{n} \rightarrow 0$ as $n\rightarrow \pm \infty$.

The current model employs complex-valued coefficients in the vertical coupling between the parallel arrays, while the previous work~\cite{kirikchi2016bright} and~\cite{kirikchi2018solitons} adopted purely imaginary and real-valued vertical coupling between the parallel arrays, respectively, which acts as the gain or loss in the system. Additionally, they also included the real-valued and purely imaginary phase-velocity mismatch between the horizontal cores in~\cite{kirikchi2016bright} and~\cite{kirikchi2018solitons}, respectively, which is absent in our current model.

For the uncoupled case, also known as the anticontinuum limit, i.e., when $\epsilon = 0$, the governing equations~\eqref{1model} reduce to another \pts-symmetric system in the presence of complex-valued coupling but in the absence of discrete Laplacian terms, which has been investigated in~\cite{karthiga2016systems}. A similar setup to our model was studied in~\cite{xu2015generalized} where the authors approached from a dynamical system point of view by incorporating the so-called Stokes variables into the system.

We substitute the following expressions for the com-plex-valued wave functions $u_n$ and $v_n$ to obtain static solutions of the governing equations~\eqref{1model}:
\begin{equation}		\label{3.1a}
u_{n} = A_{n} e^{i\omega t}, 		\qquad
v_{n} = B_{n} e^{i\omega t},
\end{equation}
where the coefficients $A_{n}$, $B_{n} \in \mathbb{C}$, and $\omega\in\mathbb{R}$ is the propagation constant. We obtain the following static equations for the \pts-symmetry dimer:
\begin{equation}		\label{2static}
\begin{split}
\omega A_{n} &= |A_{n}|^{2}A_{n}+\epsilon(A_{n+1}-2A_{n}+A_{n-1})-i\gamma B_{n}+B_{n},\\ 
\omega B_{n} &= |B_{n}|^{2}B_{n}+\epsilon(B_{n+1}-2B_{n}+B_{n-1})+i\gamma A_{n}+A_{n}.
\end{split}	
\end{equation}

The static equations~\eqref{2static} for $\epsilon = 0$ has been analyzed in details in~\cite{karthiga2016systems,jorgensen1993modified,jorgensen1994hamiltonian}. For sufficiently small but nonzero linear horizontal coupling $\epsilon$, one can verify the existence of soliton solutions emerging from the anticontinuum limit by employing a generalization of the Implicit Function Theorem to a Banach space~\cite{zeidler1995applied,accinelli2010a}. We can adopt the existence analysis of~\cite{chernyavsky2016} to our system rather straightforwardly. In this paper, we only derive the asymptotic series of the soliton solutions and do not proceed to the theorem in more detail.

We express the complex-valued quantities $A_n$ and $B_n$ as perturbation expansions in terms of the small linear horizontal coupling $\epsilon$:
\begin{equation}		\label{expansionAB}
\begin{aligned}
A_{n} &= A_{n}^{(0)} + \epsilon A_{n}^{(1)} + \epsilon^2 A_{n}^{(2)} + \dots, \\
B_{n} &= B_{n}^{(0)} + \epsilon B_{n}^{(1)} + \epsilon^2 B_{n}^{(2)} + \dots.
\end{aligned}
\end{equation}
We substitute these expansions~\eqref{expansionAB} to the static equations~\eqref{2static} and collect the terms in successive powers of~$\epsilon$. We then obtain the following equations at $\mathcal{O}(1)$ and $\mathcal{O}(\epsilon)$, respectively:
\begin{equation}		\label{15timindord1}
\begin{aligned}
A_{n}^{(0)}(1 + i\gamma) &= B_{n}^{(0)}(\omega - B_{n}^{(0)} B_{n}^{*{(0)}}),		\\
B_{n}^{(0)}(1 - i\gamma) &= A_{n}^{(0)}(\omega - A_{n}^{(0)} A_{n}^{*{(0)}}).
\end{aligned}
\end{equation}
and
\begin{equation}		\label{15atimindordeps}
\begin{aligned}
A_{n}^{(1)}(1 + i\gamma) &= B_{n}^{(1)}(\omega - 2B_{n}^{(0)} B_{n}^{*{(0)}}) - {B_{n}^{(0)}}^2 B_{n}^{*(1)} - \Delta_{2}B_{n}^{(0)},\\
B_{n}^{(1)}(1 - i\gamma) &= A_{n}^{(1)}(\omega - 2A_{n}^{(0)} A_{n}^{*{(0)}}) - {A_{n}^{(0)}}^2 A_{n}^{*(1)} - \Delta_{2}A_{n}^{(0)}.
\end{aligned}
\end{equation}

The corresponding bright discrete soliton solutions admit two natural, fundamental modes for any $\epsilon > 0$, which range from the anticontinuum to anticontinuum limits. They are the one-excited- and two-excited-sites, also known as the onsite and intersite bright discrete modes, respectively. The remainder of our discussion will focus on these two natural fundamental modes.

% Subsection 2.1
\subsection{Dimers}

In the anticontinuum limit $\epsilon \to 0$, the time-independent solution of~\eqref{2static}, i.e.,~\eqref{15timindord1}, can be written as $A_{n}^{(0)} = \tilde{a}_{0} e^{i\phi{a}}$ and $B_{n}^{(0)} = \tilde{b}_{0} e^{i\phi_{b}}$, where both amplitudes are positive real valued, i.e., $\tilde{a}_{0} > 0$ and $\tilde{b}_{0} > 0$. Solving the resulting polynomial equations for $\tilde{a}_{0}$ and $\tilde{b}_{0}$ will yield~\cite{karthiga2016systems}
\begin{align}
\tilde{a}_{0} &=  \tilde{b}_{0} = 0,\\
\tilde{a}_{0} &=  \tilde{b}_0 = \sqrt{\omega - \sqrt{1 + \gamma^2}}, 		\label{symmetrica0b0c12}\\
\tilde{a}_{0} &= -\tilde{b}_0 = \sqrt{\omega + \sqrt{1 + \gamma^2}},		\label{antisyma0b0c13}\\
\tilde{a}_{0} &=  \frac{1}{\sqrt{2}} \sqrt{\omega + \sqrt{\omega^2 - 4(1 + \gamma^2)}},	\nonumber\\
\tilde{b}_{0} &= \frac{1}{2} \frac{\sqrt{\omega + \sqrt{\omega^2 - 4(1 + \gamma^2)}} \left[\omega - \sqrt{\omega^2 - 4(1 + \gamma^2)} \right]}{\sqrt{2(1 + \gamma^2)}},	\label{asyma0b0c11}
\end{align}
and the phase $\phi_{b} - \phi_{a} = \arctan\gamma$. Due to the gauge invariance in the phase of the \pts-symmetry system~\eqref{1model}, we can set $\phi_a = 0$ without loss of generality. Thus, $\phi_{b} = \arctan(\gamma)$. Solutions~\eqref{symmetrica0b0c12},~\eqref{antisyma0b0c13}, and~\eqref{asyma0b0c11} are referred to as the symmetric, antisymmetric, and asymmetric solutions, respectively. The asymmetric solution \eqref{asyma0b0c11} emanates from a pitchfork bifurcation from the symmetric solution~\eqref{symmetrica0b0c12} at $\omega = 2\sqrt{1 + \gamma^2}$.

Another variant of interesting dimers where the coupling between the oscillators provide gain to the system was considered in \cite{alex14,dana15,kirikchi2018solitons}. Such a system may model the propagation of electromagnetic waves in coupled waveguides embedded in an active medium. The dimer considered herein when $\epsilon \to 0$ is different as in our case the coupling between the cores does not only provide gain but also loss. 

% Subsection 2.2
\subsection{Intersite discrete solitons}

The mode structure of the intersite discrete solitons in the anticontinuum limit is given by
\begin{equation}	\label{An}
\begin{aligned}
A_{n}^{(0)} &= \left\{
\begin{array}{ll}
\tilde{a}_{0} & \quad n = 0, 1,\\
0			  & \quad \text{otherwise},
\end{array}\right. \\ 
B_{n}^{(0)} &= \left\{
\begin{array}{ll}
\tilde{b}_{0}e^{i\phi_{b}}  & \quad n = 0,1,\\
0							& \quad \text{otherwise}.
\end{array}\right.
\end{aligned}
\end{equation}
For $\epsilon \neq 0$, we can write the first-order correction coefficients as $A_{n}^{(1)} = \tilde{a_{1}}$ and $B_{n}^{(1)} = \tilde{b_{1}}e^{i\phi_{b}}$. Substituting these to the $\mathcal{O}(\epsilon)$ equations~\eqref{15atimindordeps} gives the following expressions for $\tilde{a}_1$ and $\tilde{b}_1$:
\begin{equation}
\begin{aligned}
\tilde{a}_{1} &= \frac{\tilde{b_{1}}(\omega - 3\tilde{b_{0}^2}) + \tilde{b_{0}}}{\sqrt{1 + \gamma^2}}, \\%	\qquad 
\tilde{b}_{1} &= \frac{\tilde{a_{1}}(\omega - 3\tilde{a_{0}^2}) + \tilde{a_{0}}}{\sqrt{1 + \gamma^2}}.			\label{a1b10}
\end{aligned}
\end{equation}
Equations~\eqref{An} and~\eqref{a1b10} give the asymptotic expansions for $A_n$ and $B_n$ up to the first-order correction for the intersite discrete solitons. Higher-order corrections can be calculated by continuing a similar procedure. Since the first two terms are sufficient for our analysis, we exclude those higher-order terms.

% Subsection 2.3
\subsection{Onsite discrete solitons}

The onsite discrete soliton in the anticontinuum limit admits the following mode structure:
\begin{equation}	\label{An2}
\begin{aligned}
A_{n}^{(0)} &= \left\{
\begin{array}{ll}
\tilde{a}_{0} & \qquad n = 0,\\
0			  & \qquad \text{otherwise},
\end{array}\right. \\		\qquad \qquad
B_{n}^{(0)} &= \left\{
\begin{array}{ll}
\tilde{b}_{0} e^{i\phi_{b}} & \quad n = 0,\\
0							& \quad \text{otherwise}.
\end{array}\right.
\end{aligned}
\end{equation}
Employing a similar calculation as in the intersite case, we acquire the following expressions for $\tilde{a}_1$ and $\tilde{b}_1$ corresponding to $\mathcal{O}(\epsilon)$ corrections derived from equations~\eqref{15atimindordeps}:
\begin{equation}
\begin{aligned}
\tilde{a}_{1} &= \frac{\tilde{b_{1}}(\omega - 3\tilde{b_{0}^2}) + 2\tilde{b_{0}}}{\sqrt{1 + \gamma^2}}, \\		\qquad 
\tilde{b}_{1} &= \frac{\tilde{a_{1}}(\omega - 3\tilde{a_{0}^2}) + 2\tilde{a_{0}}}{\sqrt{1 + \gamma^2}}.			\label{a1b1}
\end{aligned}
\vspace*{-0.1cm}
\end{equation}
The asymptotic expansions for $A_n$ and $B_n$ up to the first-order correction for the onsite discrete solitons are thus given by expressions~\eqref{An2} and~\eqref{a1b1}. Likewise, higher-order corrections can be obtained using a similar calculation. We also exclude higher-order terms for this case.

% Section 3
\section{Stability analysis}		\label{stability}

In the following, we consider six configurations, which are combinations of the intersite and onsite discrete solitons with the three solutions of the dimers~\eqref{symmetrica0b0c12}--\eqref{asyma0b0c11}. We will denote them by subscripts (i) and (o) for intersite and onsite discrete solitons, and (s), (at), and (as) for the symmetric, antisymmetric, and asymmetric types of solution, respectively. 

We analyze the linear stability of the discrete soliton solutions by solving the corresponding eigenvalue problem.
We propose a linearization ansatz for the complex-valued functions $u_n$ and $v_n$ with $|\widetilde{\epsilon}| \ll 1$, written as follows:
\begin{align*}
u_{n} &= (A_{n} + \widetilde{\epsilon}(K_{n} + i L_{n}) e^{\lambda t}) e^{i\omega t} \\
v_{n} &= (B_{n} + \widetilde{\epsilon}(P_{n} + i Q_{n}) e^{\lambda t}) e^{i\omega t}.
\end{align*}
Substituting these expressions to the governing equations~\eqref{1model}, we obtain the linearized equation at $\mathcal{O}(\widetilde{\epsilon})$:
\begin{equation}
\begin{aligned}		\label{8eigen}
\lambda{K_{n}} &=-(A_{n}^{2}-\omega)L_{n}-\epsilon(L_{n+1}-2L_{n}+L_{n-1}) + \gamma P_{n}-Q_{n},  \\
\lambda{L_{n}} &= (3A_{n}^{2}-\omega)K_{n}+\epsilon(K_{n+1}-2K_{n}+K_{n-1}) + \gamma Q_{n}+P_{n}, \\
\lambda{P_{n}} &=-\left[\text{Re}^2(B_{n}) + 3 \, \text{Im}^2(B_{n}) - \omega\right] Q_{n} - \gamma K_{n} - L_{n} \\ & \quad - \epsilon(Q_{n + 1} - 2Q_{n} + Q_{n- 1 }) - 2 \, \text{Re}(B_{n}) \, \text{Im}(B_{n}) P_{n}, \\
\lambda{Q_{n}} &= (3 \, \text{Re}^2(B_{n})+ \text{Im}^2(B_{n}) - \omega) P_{n} - \gamma L_{n} + K_{n} \\ & \quad + \epsilon(P_{n + 1} - 2P_{n} + P_{n - 1}) + 2 \, \text{Re}(B_{n}) \, \text{Im}(B_{n}) Q_{n},
\end{aligned}
\end{equation}
which need to be solved for the eigenvalue (or, spectrum) $\lambda$ and the corresponding eigenvector
\[ [\{K_{n}\}, \{L_{n}\}, \{P_{n}\}, \{Q_{n}\}]^{T}. \]
The solution $u_{n}$ is said to be (linearly) stable when Re$(\lambda) \leq 0$ for all the spectra $\lambda \in \mathbb{C}$ and unstable otherwise. However, as the spectra will come in pairs, a solution is therefore neutrally stable when Re$(\lambda) = 0$ for all $\lambda \in \mathbb{C}$.

% Subsection 3.1
\subsection{Continuous spectrum}

The eigenvalues of~\eqref{8eigen} consist of both continuous and discrete spectra. In this subsection, we investigate the former by considering the limit $n \to \pm \infty$. We introduce the following plane-wave ansatz to the eigenvector components: $K_{n} = \hat{k} e^{ikn}$, $L_{n} = \hat{l} e^{ikn}$, $P_{n} = \hat{p} e^{ikn}$, and $Q_{n} = \hat{q} e^{ikn}$, where $k \in \mathbb{R}$. Substituting these ansatzes to~\eqref{8eigen}, we obtain the following system of linear equations written in matrix form:
\begin{equation} 	\label{eigmat}
\lambda
\left[{\begin{array}{c}  
\hat{k} \\
\hat{l} \\
\hat{p} \\
\hat{q}
\end{array}}
\right] =
\left[\begin{array}{rrrr}
0    	& \xi 	 & \gamma & -1 \\
-\xi 	& 0   	 & 1      & \gamma \\
-\gamma &-1  	 & 0 	  & \xi \\
1		&-\gamma & -\xi	  &	0
\end{array} \right]
\left[{\begin{array}{c}
\hat{k} \\
\hat{l} \\
\hat{p} \\
\hat{q}
\end{array}}\right]
\end{equation}
where $\xi = \omega - 2 \epsilon(\cos k - 1)$. This matrix equation~\eqref{eigmat} can be solved analytically to yield the following linear dispersion relationship:
\begin{equation}
\lambda^2 = -(1 + \gamma^2) - \xi^2 \pm 2 |\xi| \sqrt{1 + \gamma^2}. 		\label{dispersrelat}
\end{equation}
Thus, the ranges for the continuous spectrum are given by $\lambda \in \pm[\lambda_{1{-}}, \lambda_{2{-}}]$ and $\lambda \in \pm[\lambda_{1{+}}, \lambda_{2{+}}]$, where the spectrum boundaries $\lambda_{1{\pm}}$ and $\lambda_{2{\pm}}$ lie on the imaginary axis:
\begin{align}
\lambda_{1{\pm}} &= \pm i\sqrt{1 + \gamma^2 + \omega^2 \mp 2 |\omega| \sqrt{1 + \gamma^2}},									\label{l1}\\
\lambda_{2{\pm}} &= \pm i\sqrt{1 + \gamma^2 + (\omega + 4\epsilon)^2 \mp 2|\omega + 4 \epsilon|\sqrt{1 + \gamma^2}}.  		\label{l2} 
\end{align}
We can attain these expressions~\eqref{l1} and~\eqref{l2} by substituting $k = 0$ and $k = \pi$ to the linear dispersion relationship~\eqref{dispersrelat}, respectively.

% Subsection 3.2
\subsection{Discrete spectrum}				\label{sds}

In this subsection, we seek discrete spectra of the linear eigenvalue problem~\eqref{8eigen} using a similar asymptotic expansion implemented in Section~\ref{mathmodel}, albeit with weak linear horizontal coupling $\epsilon$. Let the expansion read
\begin{equation}
X = X^{(0)} + \sqrt\epsilon X^{(1)} + \epsilon X^{(2)} + \dots,			\label{box}
\end{equation}
where $X = \left\{ \lambda, K_n, L_n, P_n, Q_n \right\}$. Then, substituting the expansion~\eqref{box} to the linear eigenvalue problem~\eqref{8eigen}, we obtain other sets of linear equations according to the successive order of $\epsilon$: $\mathcal{O}(1)$, $\mathcal{O}(\sqrt{\epsilon})$, $\mathcal{O}(\epsilon)$, etc.

At the lowest order, we attain the corresponding stability equation for the \pts-symmetric chain of dimers, in which for general values of $\gamma$ it has been elaborated in~\cite{karthiga2016systems}. Although the resulting eigenvalues have relatively simple expressions, the corresponding eigenvectors are cumbersome, which are rather worthless for further scrutiny in correcting higher-order eigenvalues. We then restraint our analysis for the case of small gain-loss parameter $\gamma$ and expand the variables~\eqref{box} in $\gamma$ for each expression at $\mathcal{O}(\epsilon^{n/2})$, $n \in \mathbb{N}_0$, obtained from~\eqref{8eigen}. We write the following:
\begin{equation}
X^{(j)} = X^{(j,0)} + \gamma X^{(j,1)} + \gamma^2 X^{(j,2)} + \dots, \nonumber
\end{equation}
where $j = 0, 1, 2, \dots$. These two small parameters $\epsilon$ and $\gamma$ are independent of each other. The steps for calculating eigenvalues $\lambda^{(j,k)}$, $j, k \in \mathbb{N}_0$ have been outlined in detail in the Appendix of~\cite{kirikchi2016bright}. Here, we will present the results instantaneously.
% Figure 1
\begin{figure*}[bthp!]
	\centering
	\includegraphics[width = 0.35\textwidth]{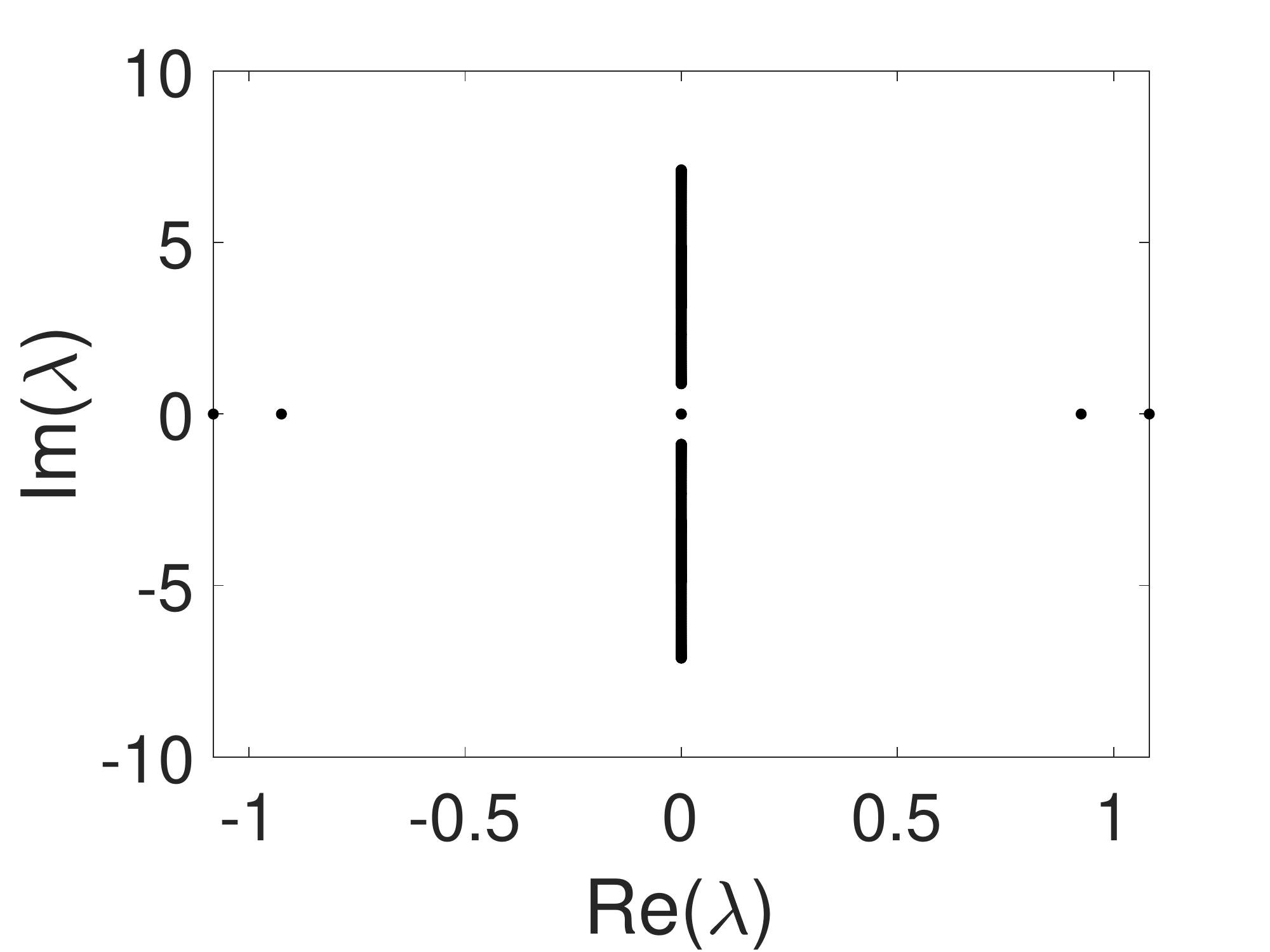} 	\hspace{1cm}
	\includegraphics[width = 0.35\textwidth]{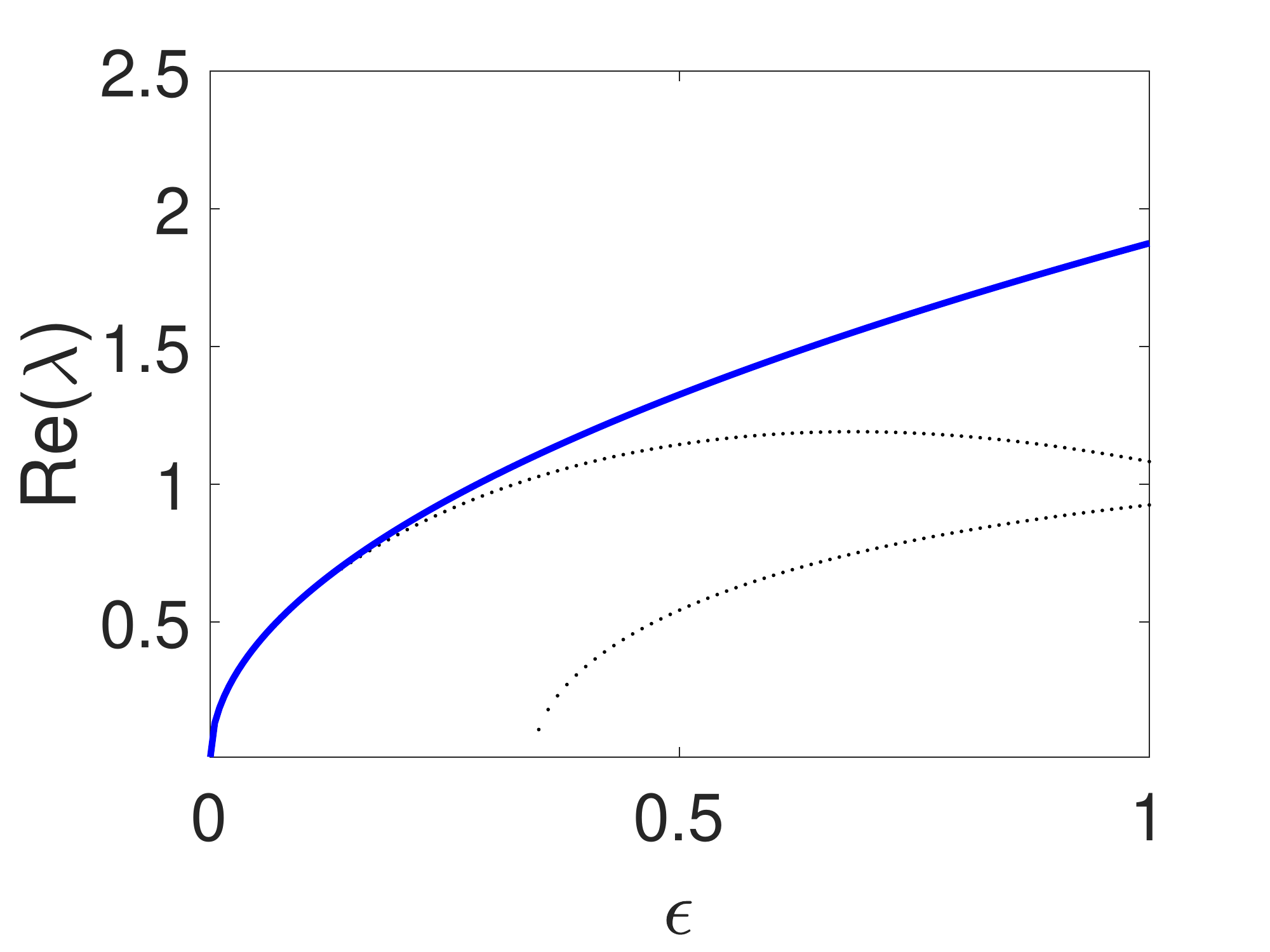}		\\
	\includegraphics[width = 0.35\textwidth]{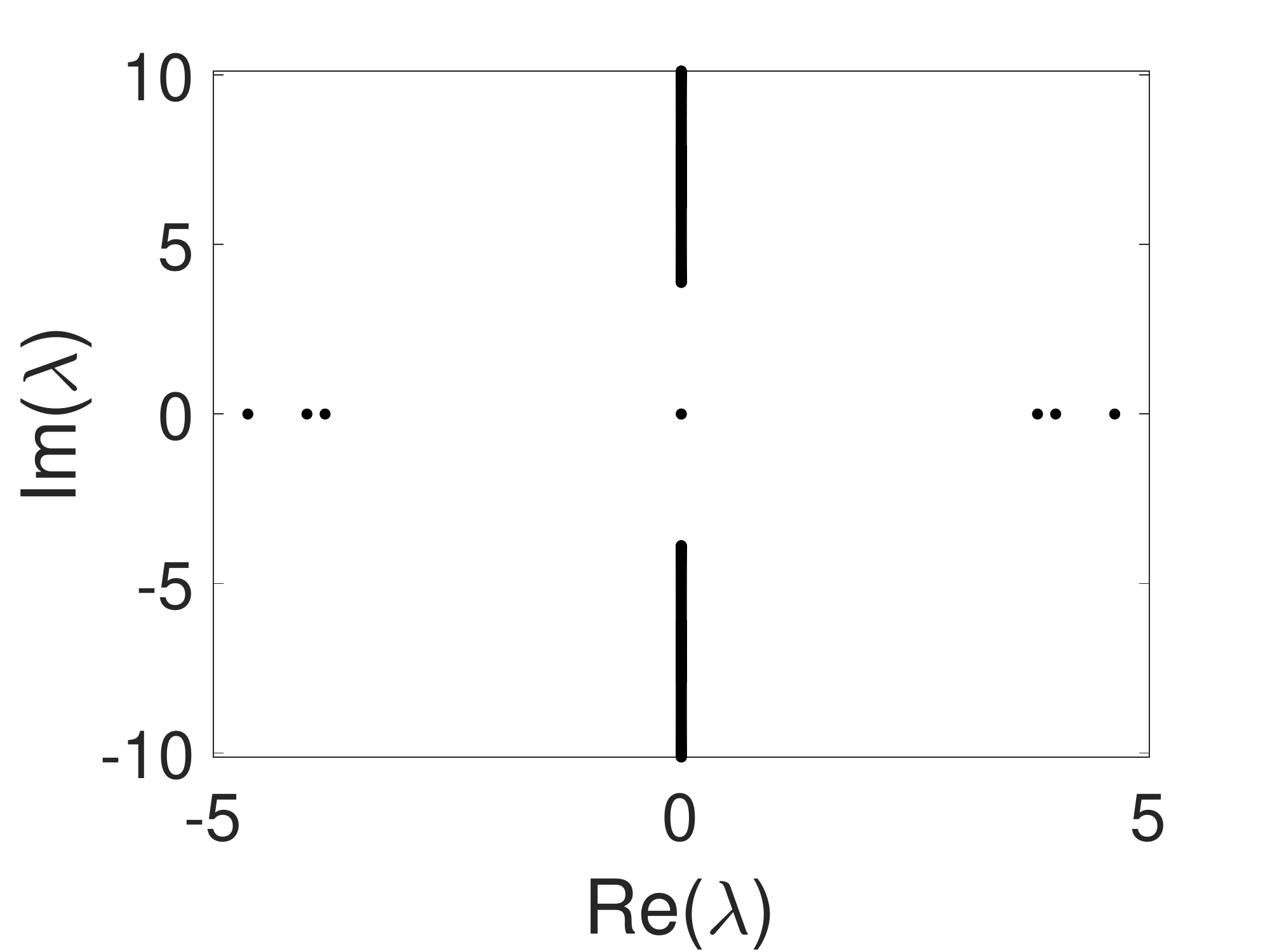} 	\hspace{1cm}
	\includegraphics[width = 0.35\textwidth]{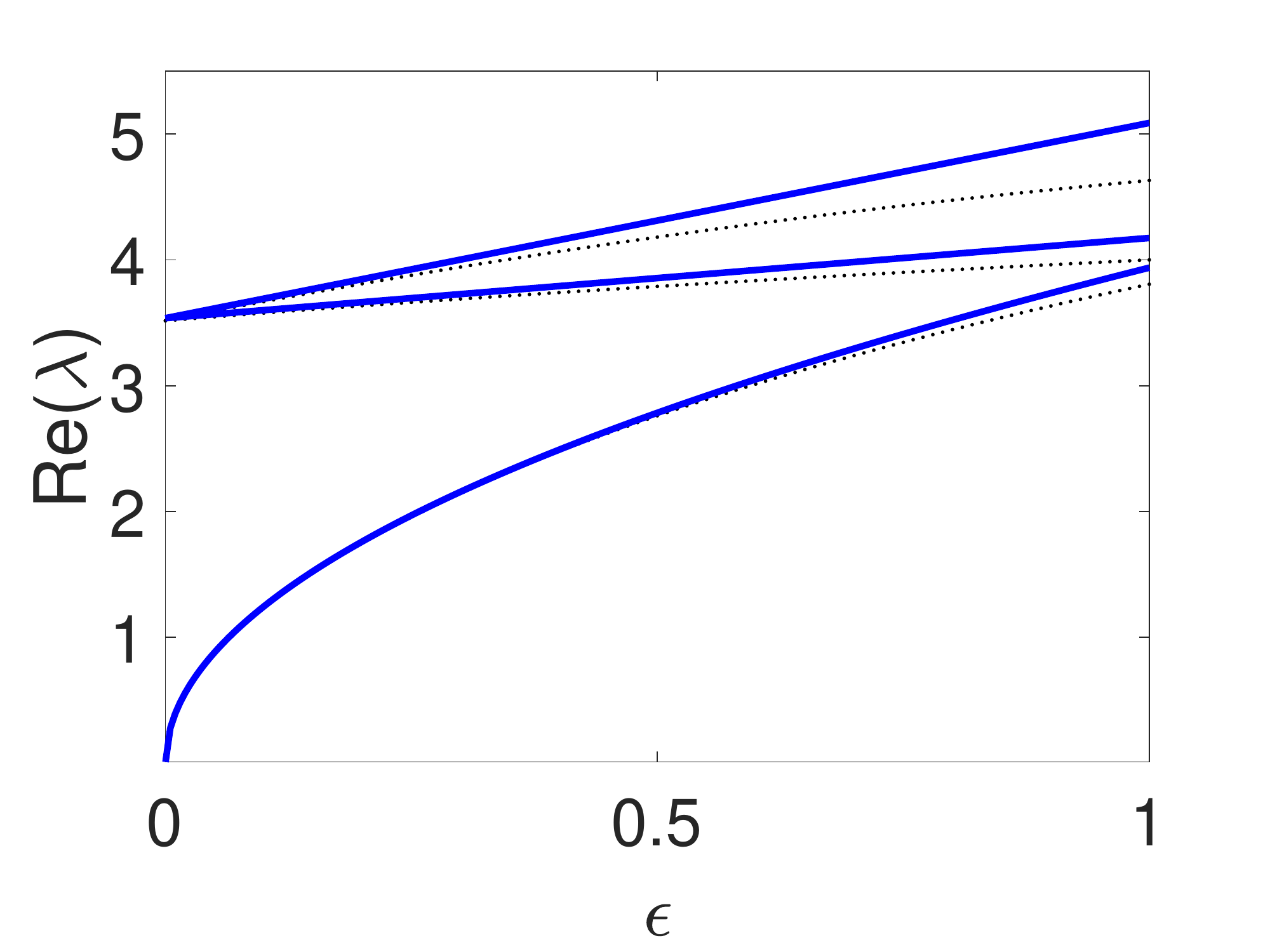}
	\caption{The spectra of unstable symmetric intersite discrete solitons with $\omega = 2$, $\gamma = 0.5$ (top panels) and $\omega = 5$, $\gamma = 0.9$ (bottom panels). The left panels feature the spectrum characteristics in the complex plane for $\epsilon = 1$. The right panels present the real-part of the spectrum as a function of the linear horizontal coupling constant~$\epsilon$. The solid blue curves are the asymptotic approximations presented in Subsubsection~\ref{is1} while the dots are obtained from a numerical calculation.} \label{fig.converge1}
\end{figure*}
% Figure 2
\begin{figure*}[bhtp!]
	\centering
	\subfloat[]{\includegraphics[width = 0.325\textwidth]{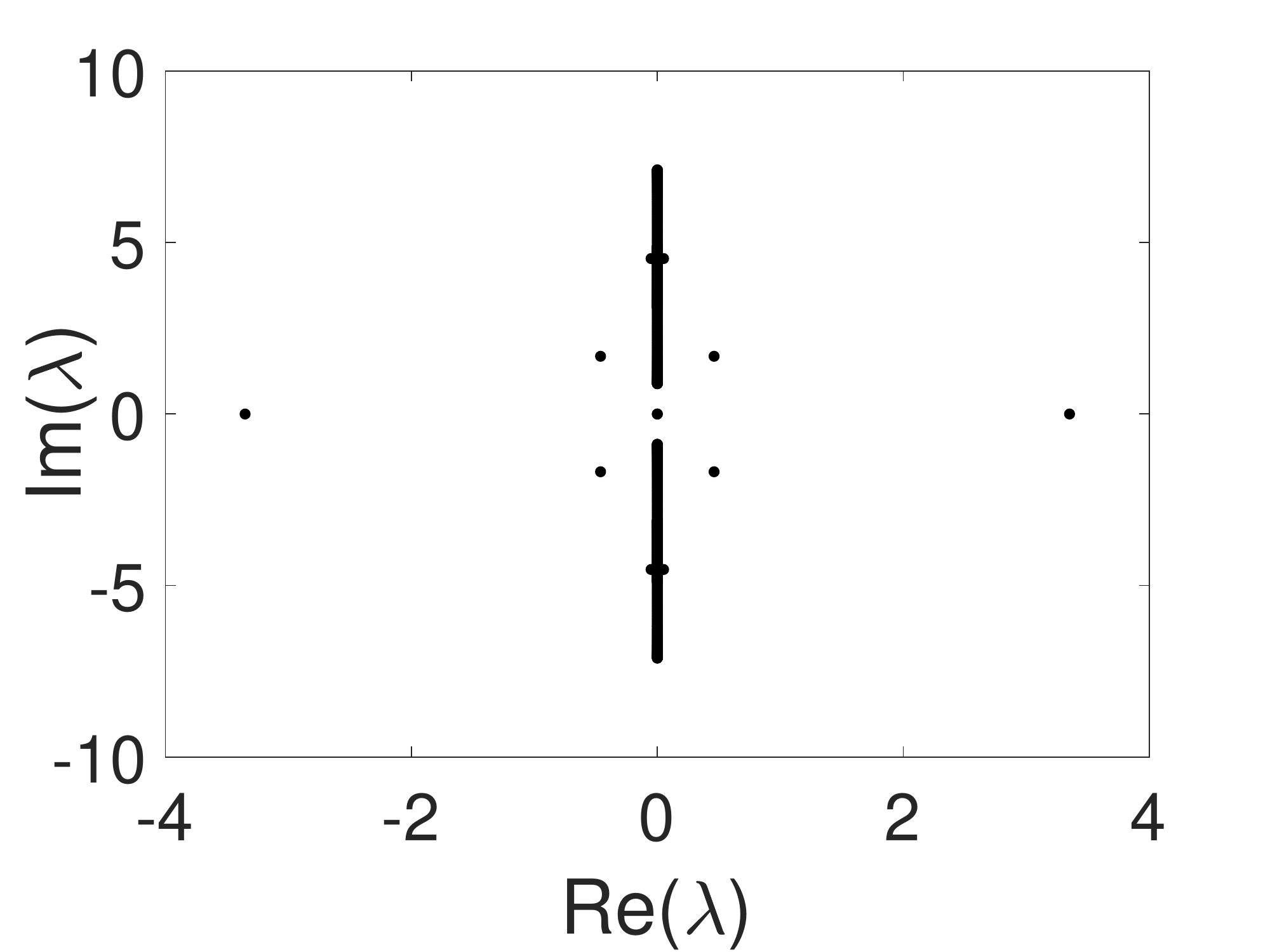}}% %\hspace{0.1cm}
	\subfloat[]{\includegraphics[width = 0.325\textwidth]{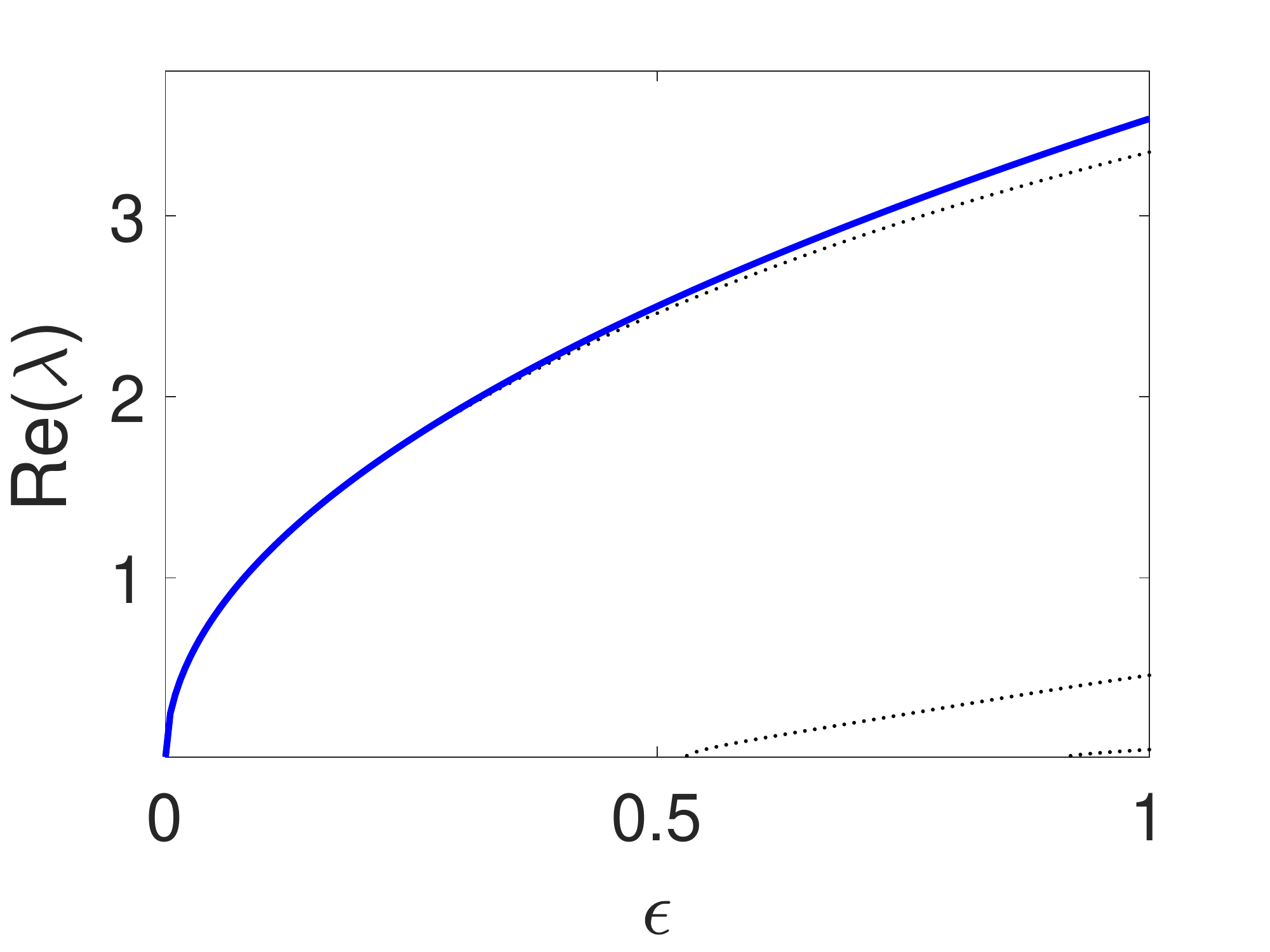}}%  %\hspace{0.1cm}
	\subfloat[]{\includegraphics[width = 0.325\textwidth]{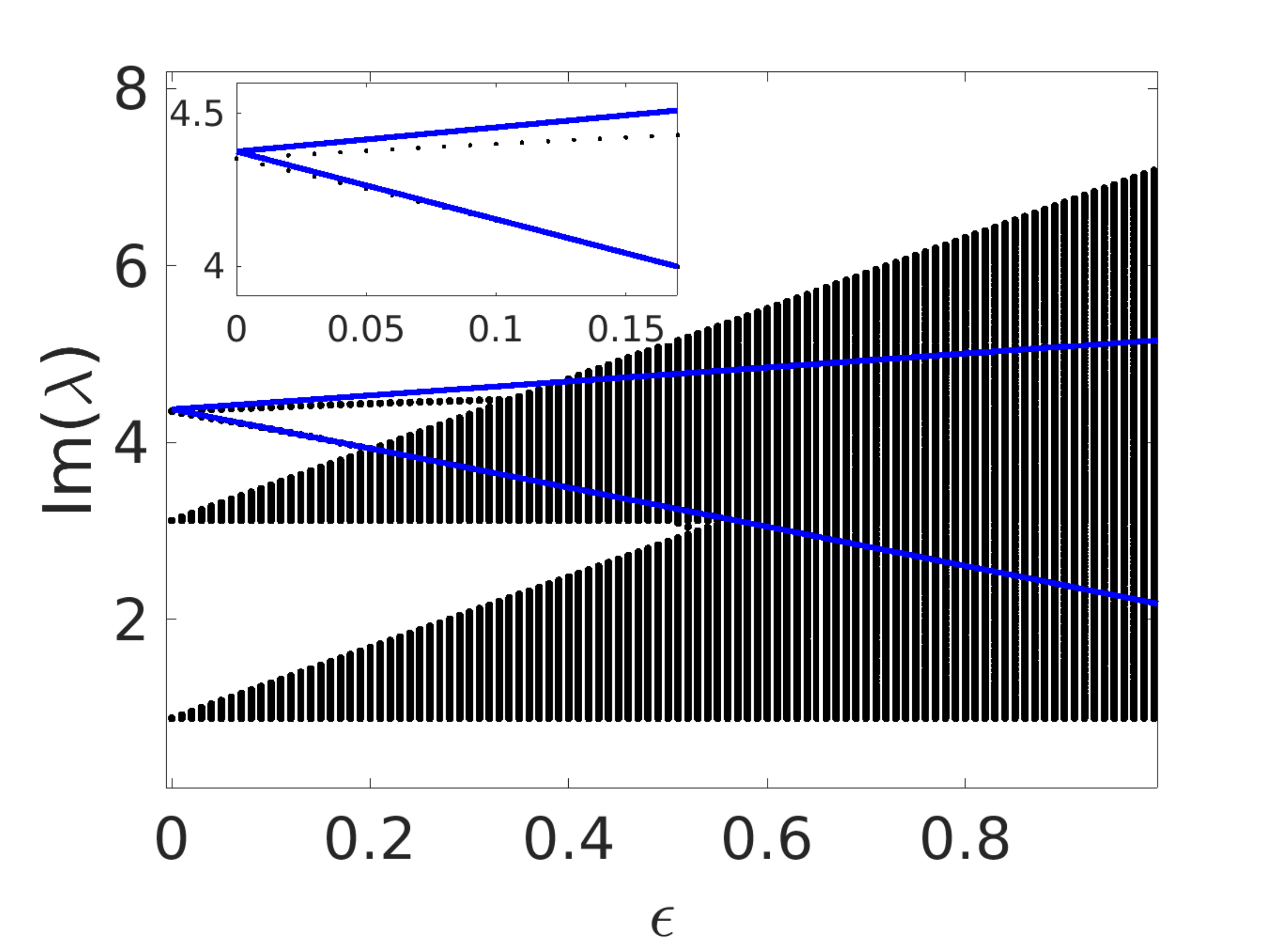}}%
	\caption{The spectra of unstable antisymmetric intersite discrete solitons with $\omega = 2$ and $\gamma = 0.5$. Panel (a) displays the spectra in the complex plane for $\epsilon = 1$. Panels (b) and (c) present the real-part and imaginary-part of the spectrum $\lambda$ as a function of the horizontal linear coupling constant $\epsilon$, respectively. The solid blue and dotted curves are attained from the asymptotic approximation and numerical calculation, respectively.} \label{fig.converge2} 
\end{figure*}

% Subsubsection 3.2.1
\subsubsection{Intersite discrete soliton}		\label{is1}			

Instead of two types of intersite discrete solitons that emerged from the analysis in~\cite{kirikchi2016bright}, i.e., symmetric and antisymmetric, we obtain an additional type, i.e., asymmetric one. All of them have in general one pair of eigenvalues that bifurcate from the origin for small $\epsilon$ and two pairs of nonzero eigenvalues. They are asymptotically given by 
\begin{align}
\lambda_{\text{(i,s)}}  &= \sqrt{\epsilon} \left(2\sqrt{\omega - 1} - \gamma^2/(2\sqrt{\omega - 1}) + \dots \right) + \mathcal{O}(\epsilon), \\ 
\lambda_{\text{(i,at)}} &= \sqrt{\epsilon} \left(2\sqrt{\omega + 1} + \gamma^2/(2\sqrt{\omega + 1}) + \dots \right) + \mathcal{O}(\epsilon), \\ 
\lambda_{\text{(i,as)}} &= \sqrt{\epsilon }\left(2\sqrt{\omega    } + 								 \dots \right) + \mathcal{O}(\epsilon),  
\end{align}
for the eigenvalues bifurcating from the origin and
\begin{align}
\lambda_{\text{(i,s)}} &= \left\{
\begin{array}{l}
\left(2\sqrt{\omega - 2} + \gamma^{2} \frac{\omega - 4}{2\sqrt{\omega - 2}} + \dots \right) \\ + \epsilon \left(\sqrt{\omega - 2} - \gamma^{2} \frac{\omega}{4\sqrt{\omega - 2}} + \dots \right) + \mathcal{O}\left(\epsilon^{3/2} \right), \\
\left(2\sqrt{\omega - 2} + \gamma^{2} \frac{\omega - 4}{2\sqrt{\omega - 2}} + \dots \right) \\ + \epsilon \left(\frac{1}{\sqrt{\omega - 2}} + \gamma^{2} \frac{\omega}{4(\omega - 2)^{3/2}} + \dots \right) + \mathcal{O}\left(\epsilon^{3/2} \right), 
\end{array}
\right. \\
\lambda_{\text{(i,at)}} &= \left\{
\begin{array}{l}
\left(2i \sqrt{\omega + 2} + \gamma^{2} \frac{i(\omega + 4)}{2\sqrt{\omega + 2}} + \dots \right) \\ - \epsilon \left(i\sqrt{\omega + 2} + \gamma^{2} \frac{3i(\omega^2 + 5 \omega + 4)}{8(\omega + 2)^{3/2}} + \dots\right) + \mathcal{O}\left(\epsilon^{3/2} \right),\\
\left(2i \sqrt{\omega + 2} + \gamma^{2} \frac{i(\omega + 4)}{2\sqrt{\omega + 2}} + \dots \right) \\ + \epsilon\left(\frac{i}{\sqrt{\omega + 2}} + \gamma^{2} \frac{i(5\omega^2 + 21 \omega + 12)}{8(\omega + 2)^{3/2}} + \dots \right) + \mathcal{O}\left(\epsilon^{3/2} \right), 
\end{array}
\right. \\
\lambda_{\text{(i,as)}} &= \left\{
\begin{array}{l}
\left(\sqrt{4 - \omega^2} - \gamma^{2} \frac{2i}{\sqrt{\omega^2 - 4}} + \dots \right) \\ + \epsilon \left(\frac{3i\omega}{\sqrt{\omega^2 - 4}} + \gamma^{2} \frac{6i\omega}{(\omega^2 - 4)^{3/2}} + \dots \right) + \mathcal{O}\left(\epsilon^{3/2} \right), \\
\left(\sqrt{4 - \omega^2} - \gamma^{2} \frac{2i}{\sqrt{\omega^2 - 4}} + \dots \right) \\ + \epsilon \left(\frac{i\omega}{\sqrt{\omega^2 - 4}} + \gamma^{2} \frac{2i\omega}{(\omega^2 - 4)^{3/2}} + \dots \right) + \mathcal{O}\left(\epsilon^{3/2} \right), \\
\end{array}
\right.
\end{align}
for the nonzero eigenvalues. 

% Subsubsection 3.2.2
\subsubsection{Onsite discrete soliton}		 		\label{os1}

\vspace*{-0.147cm}
Similarly, we also have three types of onsite discrete solitons with each one generally has only one pair of nonzero eigenvalues.
For a small value of $\epsilon$, the symmetric, antisymmetric, and asymmetric onsite discrete solitons are given asymptotically as follows, respectively: %\\
\begin{align}
\lambda_{\text{(o,s)}} &= \left(2\sqrt{\omega - 2} + \gamma^{2} \frac{(\omega - 4)}{2\sqrt{\omega - 2}} + \dots \right) \nonumber \\ & \quad + \epsilon \left(\frac{2}{\sqrt{\omega - 2}} + \gamma^{2} \frac{\omega}{2(\omega - 2)^{3/2}} + \dots \right) + \mathcal{O}\left(\epsilon^{3/2} \right), %\\
\end{align}
\begin{align}
\lambda_{\text{(o,at)}} &= \left(2i\sqrt{\omega + 2} + \gamma^{2} \frac{i(\omega + 4)}{2\sqrt{\omega + 2}} + \dots \right) \nonumber \\ & \quad + \epsilon \left(\frac{2i}{\sqrt{\omega + 2}} + \gamma^{2} \frac{i\omega}{2(\omega + 2)^{3/2}} + \dots \right) + \mathcal{O}\left(\epsilon^{3/2} \right), \\
%\end{align}
%\begin{align}
\lambda_{\text{(o,as)}} &= \left(i\sqrt{\omega^2 - 4} - \gamma^{2} \frac{2i}{\sqrt{\omega^2 - 4}} + \dots \right) \nonumber \\ & \quad + \epsilon \left(\frac{2i\omega}{\sqrt{\omega^2 - 4}} + \gamma^{2} \frac{4i\omega}{(\omega^2 - 4)^{3/2}} + \dots\right) + \mathcal{O}\left(\epsilon^{3/2} \right).
\end{align}
% Figure 3
\begin{figure*}[hbtp!]
	\centering
	\subfloat[]{\includegraphics[width = 0.325\textwidth]{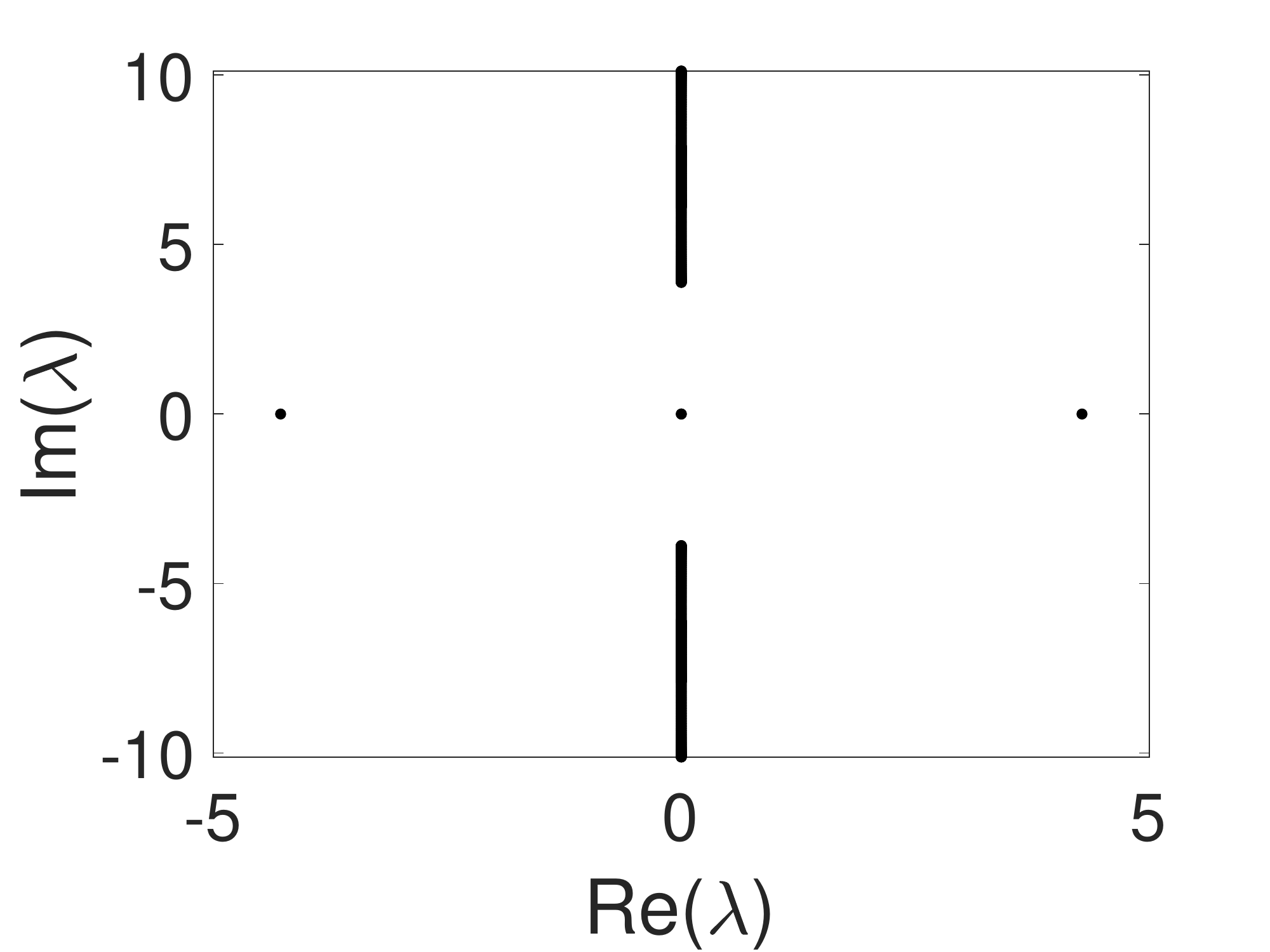}}
	\subfloat[]{\includegraphics[width = 0.325\textwidth]{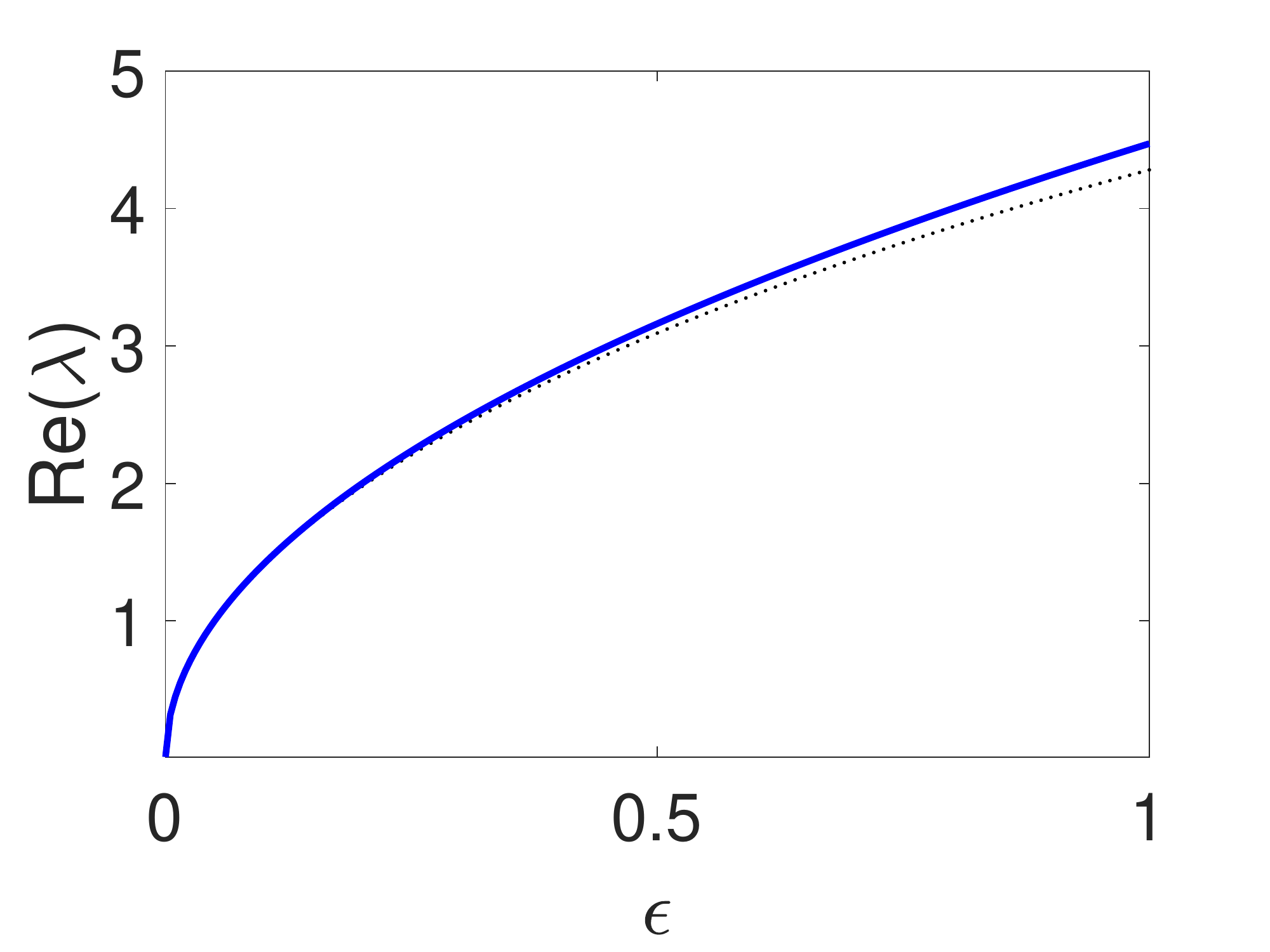}}
	\subfloat[]{\includegraphics[width = 0.325\textwidth]{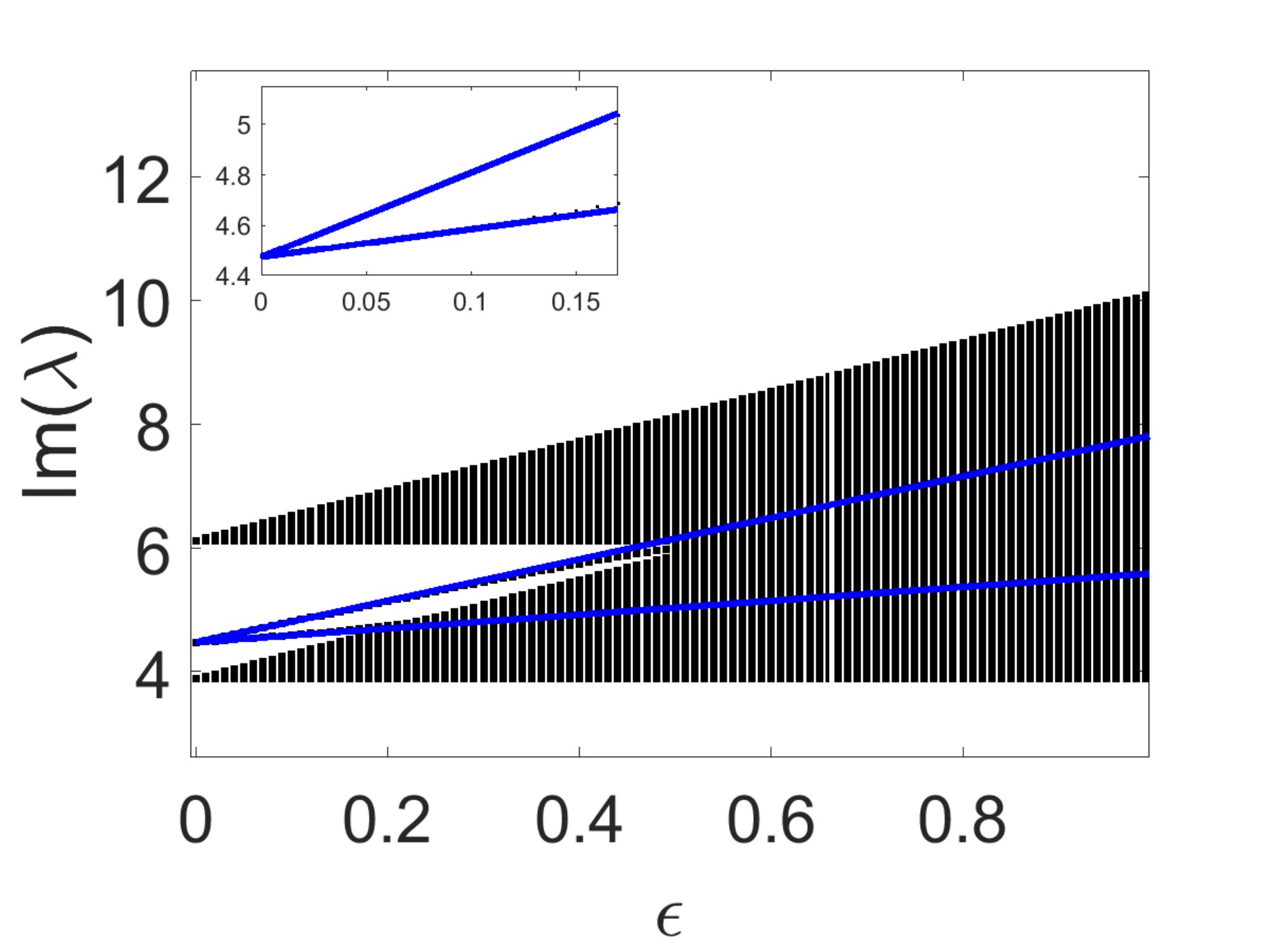}}
	\caption{The spectra of unstable asymmetric intersite discrete solitons with $\omega = 5$ and $\gamma = 0.5$. Panel (a) displays the spectra in the complex plane for $\epsilon = 1$. Panels (b) and (c) present the eigenvalues $\lambda$ as a function of the coupling constant $\epsilon$. The solid blue and dotted curves are attained from the asymptotic approximation and numerical calculation, respectively.} \label{fig.converge3}
\end{figure*}

% Section 4
\section{Numerical results}			\label{numerical}

To solve the static equations~\eqref{2static} numerically, we implemented the Newton-Raphson method. After we obtain the numerical soliton solutions, we investigate their stability by solving the corresponding linear eigenvalue problem~\eqref{8eigen} using a standard eigenvalue problem solver. In this section, we will compare the analytical calculations obtained in the previous sections with the numerical results.

First, we examine the family of symmetric intersite discrete solitons. The left panels of Figure~\ref{fig.converge1} display both continuous and discrete spectra in the complex plane for $\epsilon = 1$ and the right panels of Figure~\ref{fig.converge1} reveal the dynamics of the real-valued spectrum as a function of the linear horizontal coupling~$\epsilon$. The parameter values for the top panels of Figure~\ref{fig.converge1} are $\omega = 2$ and $\gamma = 0.5$. For these parameter values, there exists only one pair of discrete spectrum in the beginning (anticontinuum limit). As the coupling~$\epsilon$ value increases, one of the nonzero spectra that was initially on the imaginary axis becomes real-valued, too.

The bottom panels of Figure~\ref{fig.converge1} demonstrate the spectra for a sufficiently large value of the propagation constant ($\omega = 5$ and $\gamma = 0.9$). We observe that in the anticontinuum limit, one pair of the discrete spectrum is located at the origin while two pairs lie on the real axis. As the linear horizontal coupling~$\epsilon$ increases, the pair that was initially at the origin moves closer to the other two pairs. In the right panels, we also illustrate the approximate plot for the discrete spectrum in solid blue curves, where a satisfying agreement is attained for small values of the coupling~$\epsilon$. In all cases, the family of symmetric intersite discrete solitons is unstable in the continuum limit $\epsilon \to \infty$.

Second, we consider the family of antisymmetric intersite discrete solitons. Figure~\ref{fig.converge2}(a) displays a typical distribution of both continuous and discrete spectra in the complex plane for $\epsilon = 1$, $\omega = 2$ and $\gamma = 0.5$. Figures~\ref{fig.converge2}(b) and~\ref{fig.converge2}(c) feature the real-part and imaginary-part of the discrete spectrum as a function of the linear horizontal coupling parameter~$\epsilon$, respectively. We notice that there exists a pair of the discrete spectrum that bifurcates from the origin. For this particular value of the propagation constant, the spectra satisfy the condition $\lambda^2 < \lambda_{2-}^2$ in the anticontinuum limit $\epsilon \to 0$. As the linear horizontal coupling~$\epsilon$ increases, the continuous and discrete spectra collide, and consequently, the eigenvalues become complex-valued. Similar to the previous case, the family of antisymmetric intersite discrete solitons is unstable in the continuum limit $\epsilon \to \infty$ for all the chosen parameter values.

Third, the final case for intersite discrete solitons is the family of asymmetric ones. Figure~\ref{fig.converge3} displays a common spectrum distribution in the complex plane for a particular choice of parameters $\omega$ and $\gamma$. Although the complex eigenvalues are not visible, the asymmetric intersite discrete solitons yield unstable solutions for the set of calculated parameters in the continuum limit. In the anticontinuum limit, the position of the discrete spectrum for the previous case of the antisymmetric intersite is above all the continuous spectrum, \emph{viz.} Figure~\ref{fig.converge2}. The main interesting part is that the unstable eigenvalues bifurcate into the complex plane, i.e., the emergence of eigenvalues with the nonzero imaginary part. For the asymmetric intersite case, the position of the discrete spectrum is in between the continuous one and the imaginary part remains zero.

\begin{figure}[htbp!]
\centering
\subfloat[]{\includegraphics[width = 0.37\textwidth]{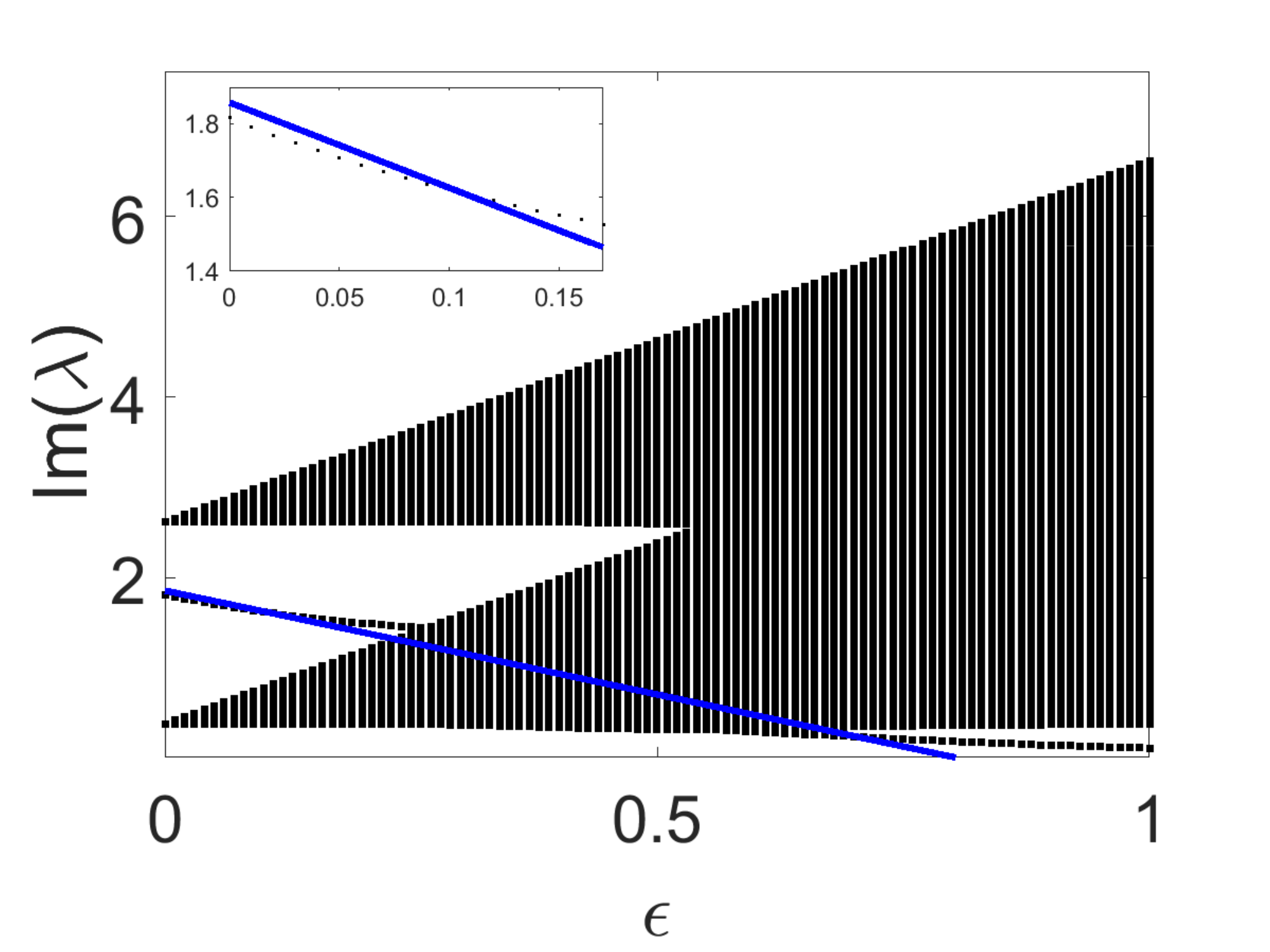}} \\ %\hspace{0.1cm} 
\subfloat[]{\includegraphics[width = 0.37\textwidth]{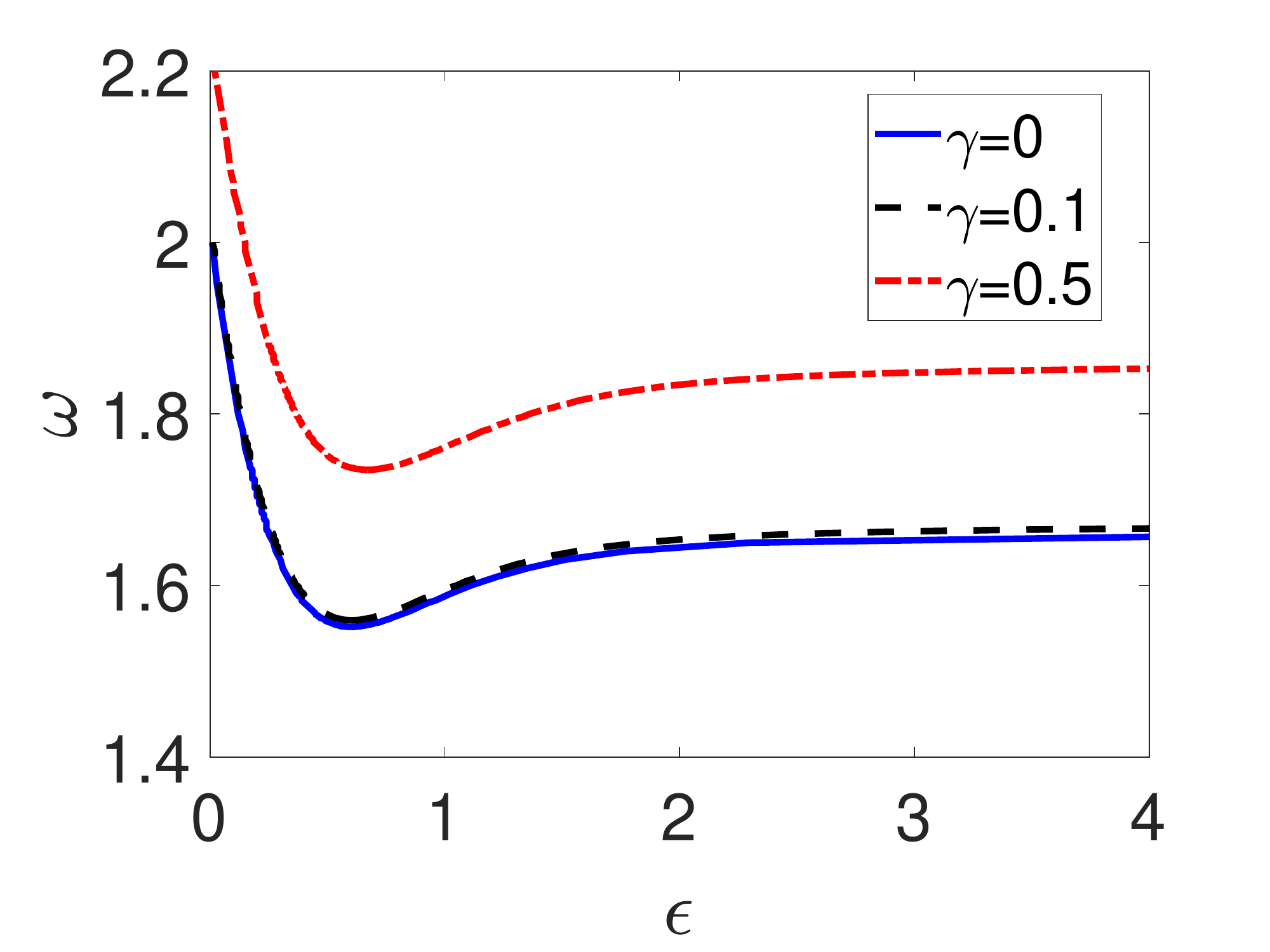}}
\caption{(a) The imaginary-part of the spectrum as a function of the linear horizontal coupling parameter~$\epsilon$ and its approximation of symmetric onsite discrete solitons with $\omega = 1.5$, $\gamma = 0.5$. (b) The stability region of the family of symmetric onsite discrete solitons in the $(\epsilon,\omega)$-plane for three distinct values of $\gamma$. The family of symmetric onsite discrete solitons are unstable above the curves.} \label{fig.converge4}
\end{figure}

\begin{figure*}[htbp!]
\centering
\subfloat[]{\includegraphics[width = 0.325\textwidth]{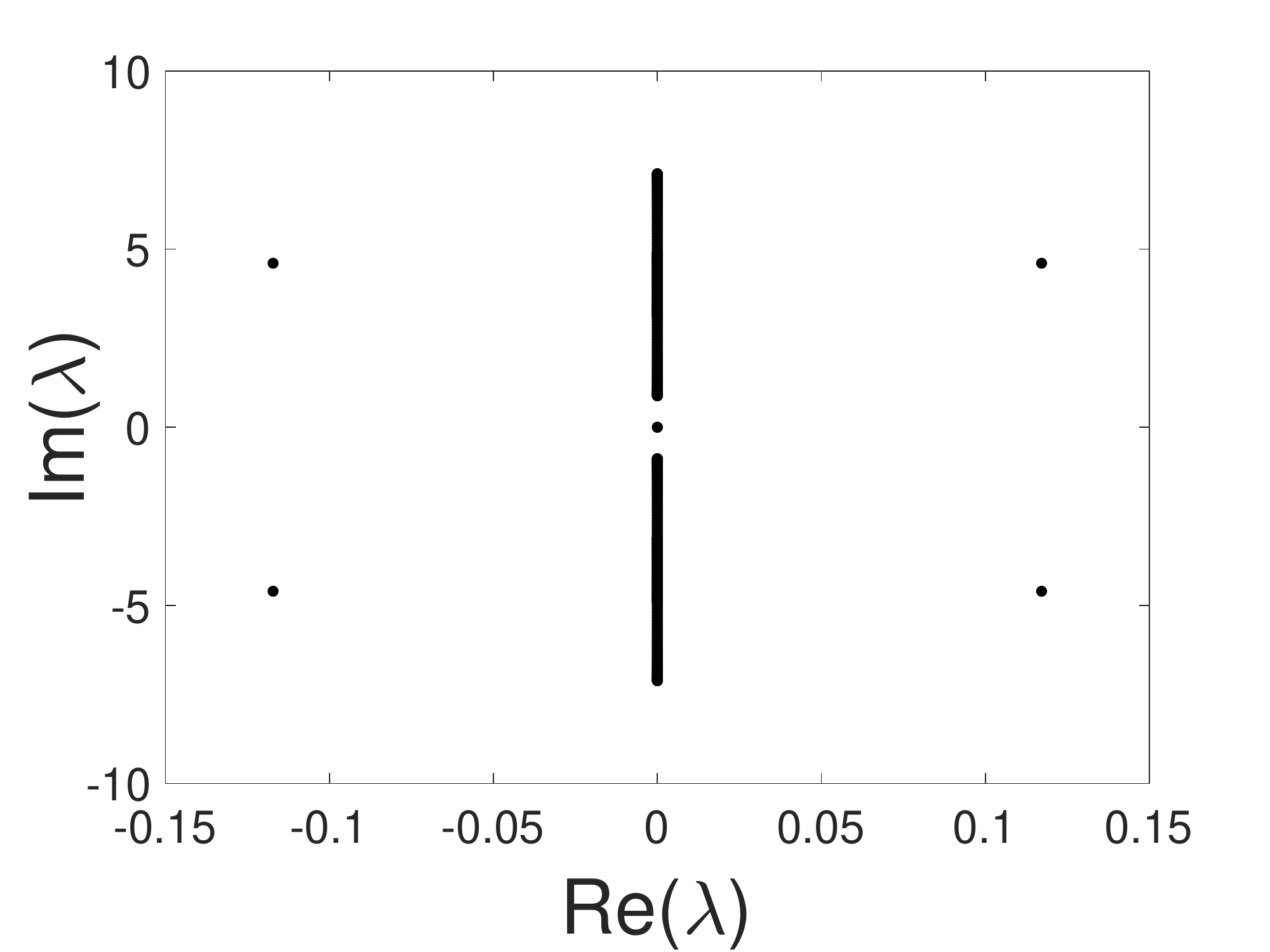}}
\subfloat[]{\includegraphics[width = 0.325\textwidth]{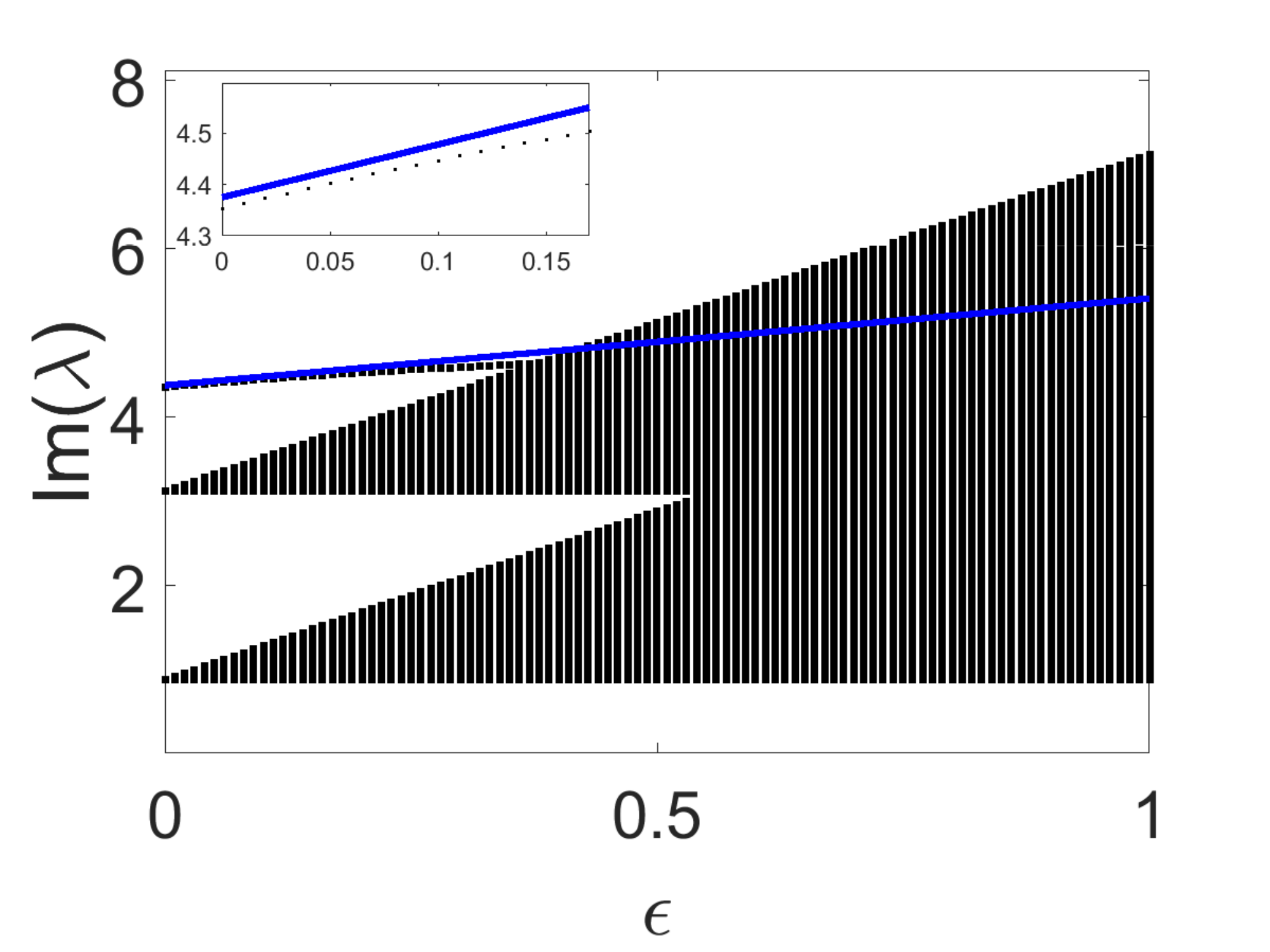}}
\subfloat[]{\includegraphics[width = 0.325\textwidth]{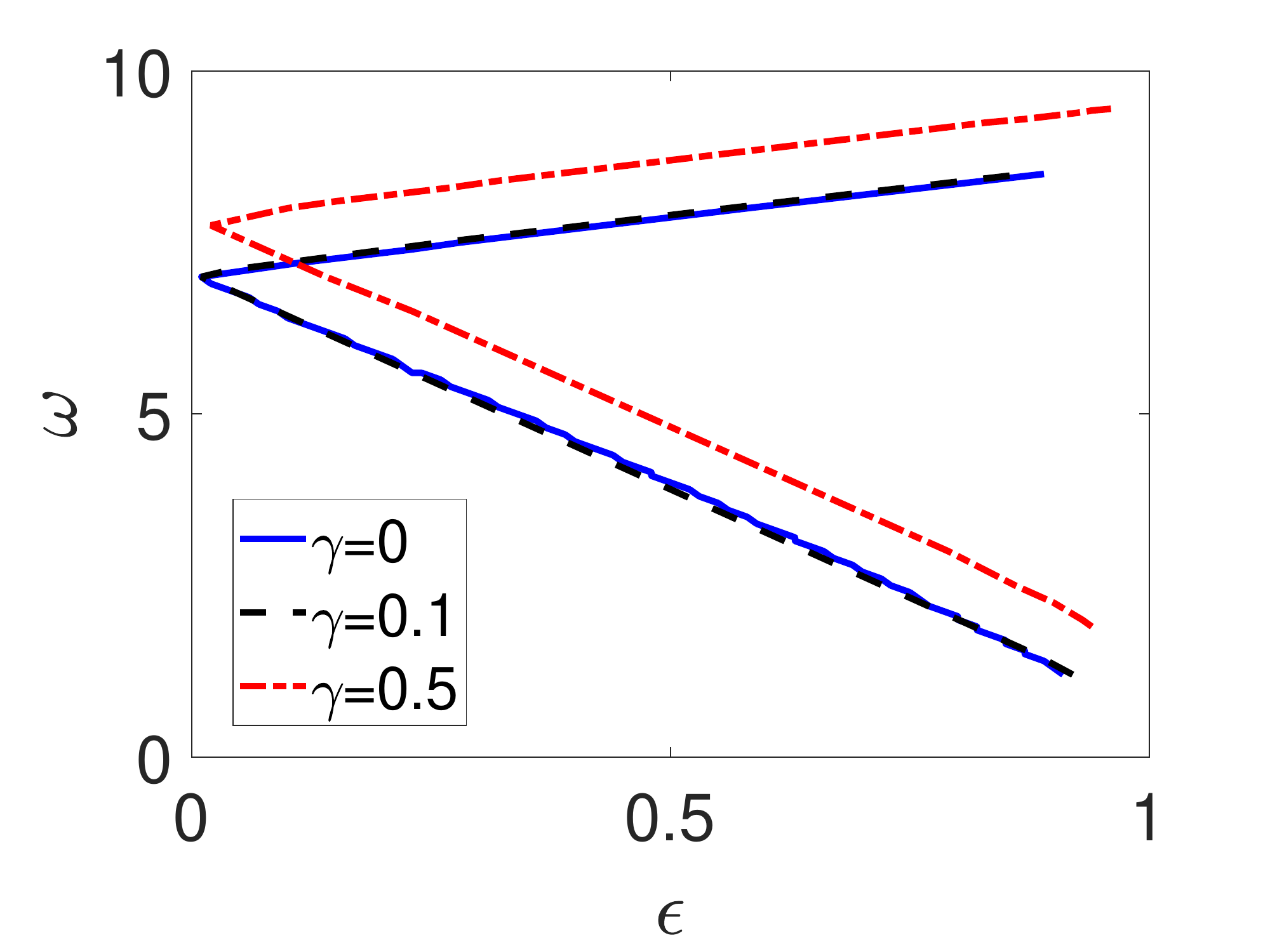}}
\caption{Panel (a) displays eigenvalues of antisymmetric onsite discrete soliton for $\omega = 2$, $\gamma = 0.5$, and $\epsilon = 1$. (b) The imaginary-part of the spectrum as a function of the linear horizontal coupling parameter~$\epsilon$. (c) The stability diagram of the discrete solitons for several values of $\gamma$. The family of antisymmetric onsite discrete solitons is unstable between the curves.} \label{fig.converge5} 
\end{figure*}

\begin{figure*}[htbp!]
\centering
\includegraphics[width = 0.35\textwidth]{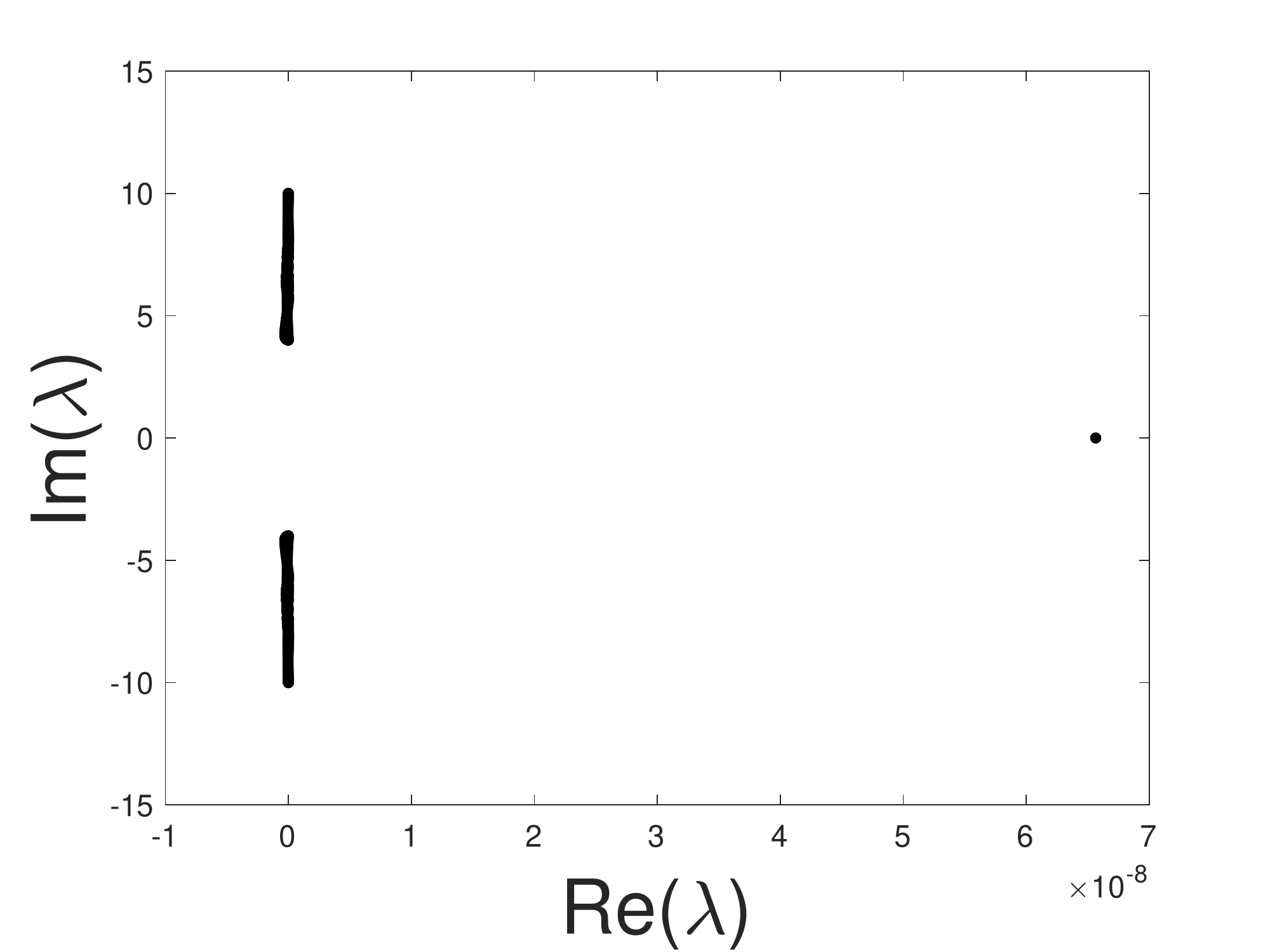} 		\hspace{1cm}
\includegraphics[width = 0.35\textwidth]{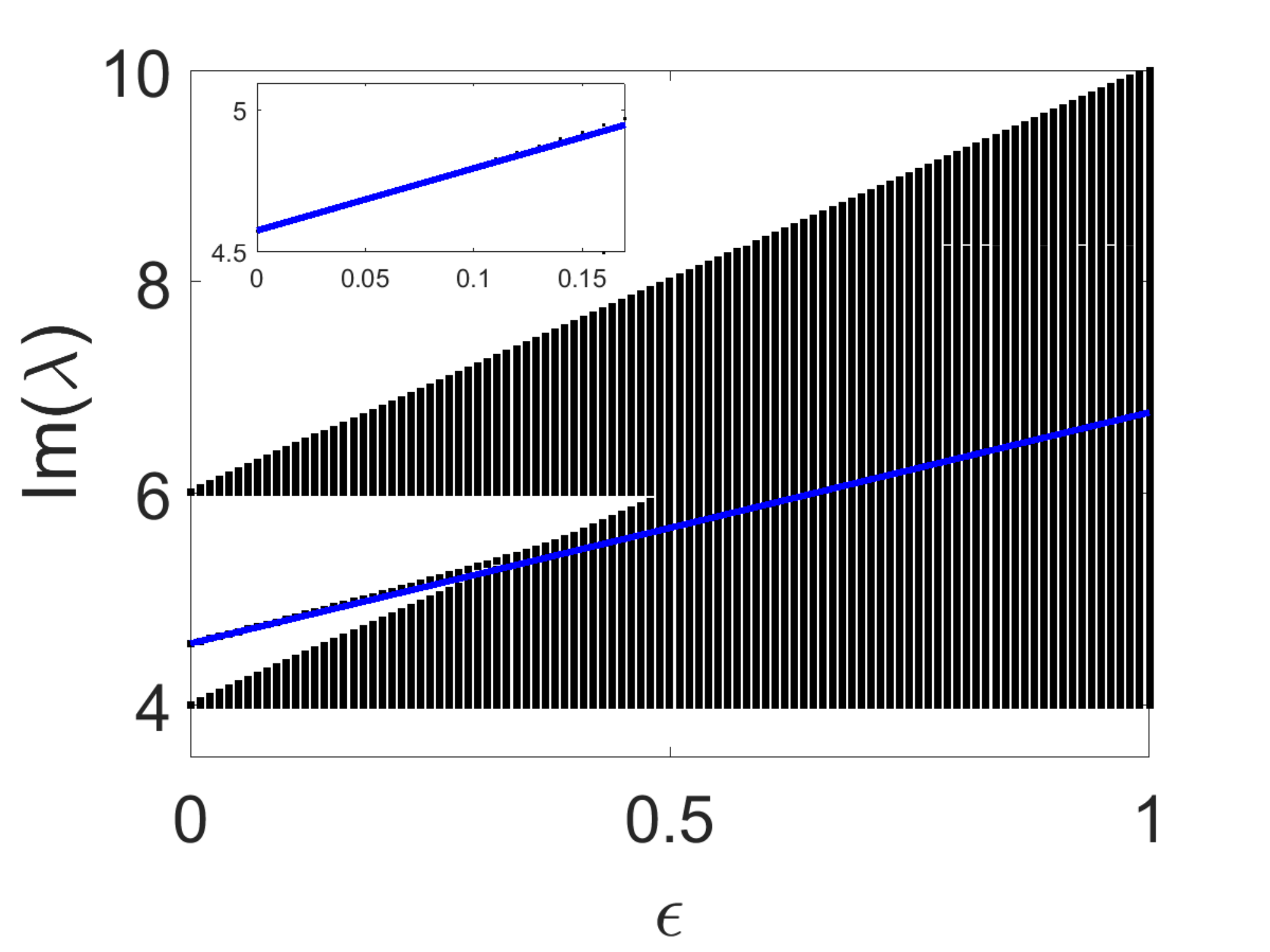}
\caption{The left panel displays the eigenvalues of asymmetric onsite discrete solitons for $\omega = 5$, $\gamma = 0.1$, and $\epsilon = 1$. The right panel shows the imaginary-part of the spectrum as a function of the linear horizontal coupling parameter~$\epsilon$.} \label{fig.converge6}
\end{figure*}

Figure~\ref{fig.converge4}--\ref{fig.converge6} present the spectrum characteristics for the families of onsite discrete solitons. Different from the families of intersite discrete solitons that are consistently unstable, the families of onsite discrete solitons can be stable depending on the parameter values. Figure~\ref{fig.converge4}(a) displays the imaginary-part of the discrete spectrum as a function of the linear horizontal coupling coefficient~$\epsilon$ for the family of symmetric onsite discrete solitons. The choice of $\omega$, in this case, corresponds to stable discrete solitons. However, there are regions of instability for different parameter values of $\omega$ that may depend on $\gamma$ and $\epsilon$. Figure~\ref{fig.converge4}(b) displays the regions of (in)stability for the symmetric onsite discrete solitons in the $(\epsilon, \omega)$-plane for three distinct values of $\gamma$. We observe that the real-part of the contribution of the gain-loss parameter~$\gamma$ toward the \pts-symmetric system for this particular case seems to be beneficial since it expands the region of stability for the family of symmetric onsite discrete solitons.

Figure~\ref{fig.converge5} shows a typical feature of the spectrum for the family of antisymmetric onsite discrete solitons. As depicted in Figure~\ref{fig.converge5}(a) for $\epsilon = 1$, due to the presence of quartet complex-valued eigenvalues, this family of solitons is generally unstable. Since the instability occurs due to the collision of the discrete spectrum with the continuous one, stability regions may present before the encounter. Figure~\ref{fig.converge5}(c) shows the sectors where the family of antisymmetric onsite discrete solitons is unstable between the curves. These solitons are unstable in the continuum limit $\epsilon \to \infty$. Figure~\ref{fig.converge6} shows the family of asymmetric onsite discrete solitons that is stable in the region of their existence. Note that this family of solitons bifurcates from the symmetric ones.

\begin{figure*}[htbp!]
\centering
\includegraphics[width = 0.35\textwidth]{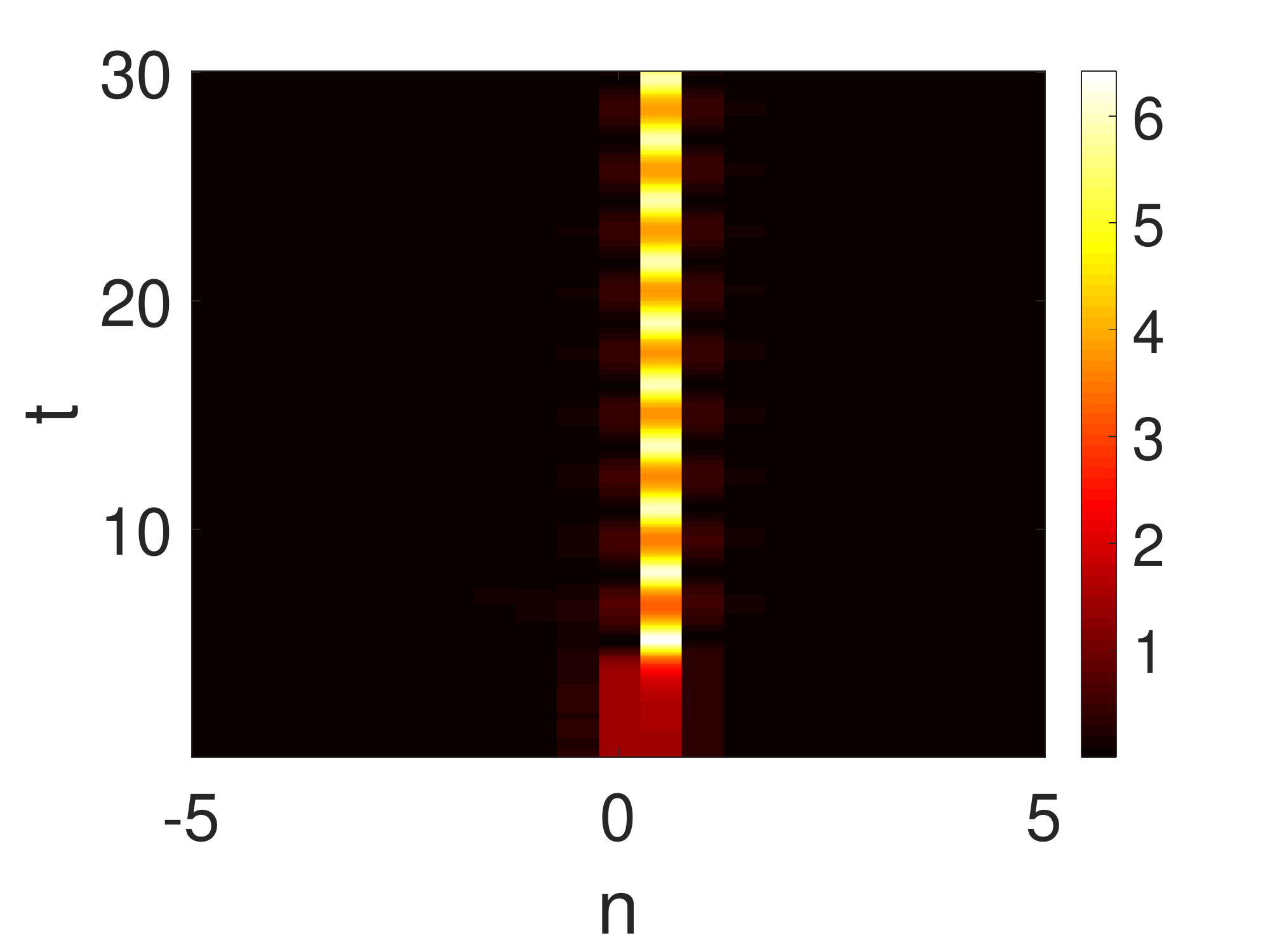}			\hspace{1cm}
\includegraphics[width = 0.35\textwidth]{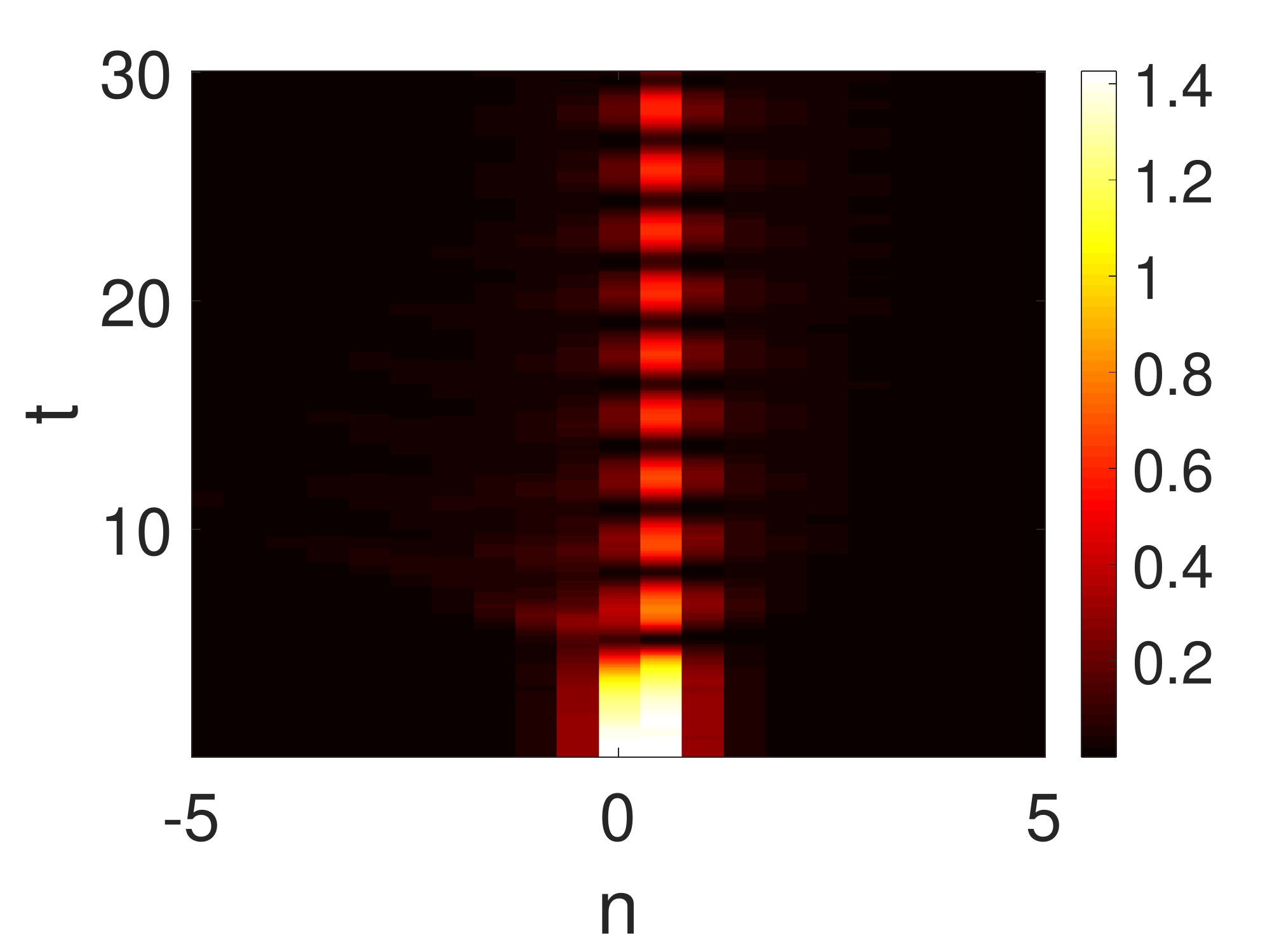}
\caption{The typical dynamics of unstable symmetric intersite discrete solitons with $\omega = 2$, $\gamma = 0.5$, $\epsilon = 1$ (compare with Figure~\ref{fig.converge1}). The left and right panels depict the plots for $|u_n|^2$ and $|v_n|^2$, respectively.} 		\label{fig.converge7}
\end{figure*}

\begin{figure*}[tbhp!]
\centering
\includegraphics[width = 0.35\textwidth]{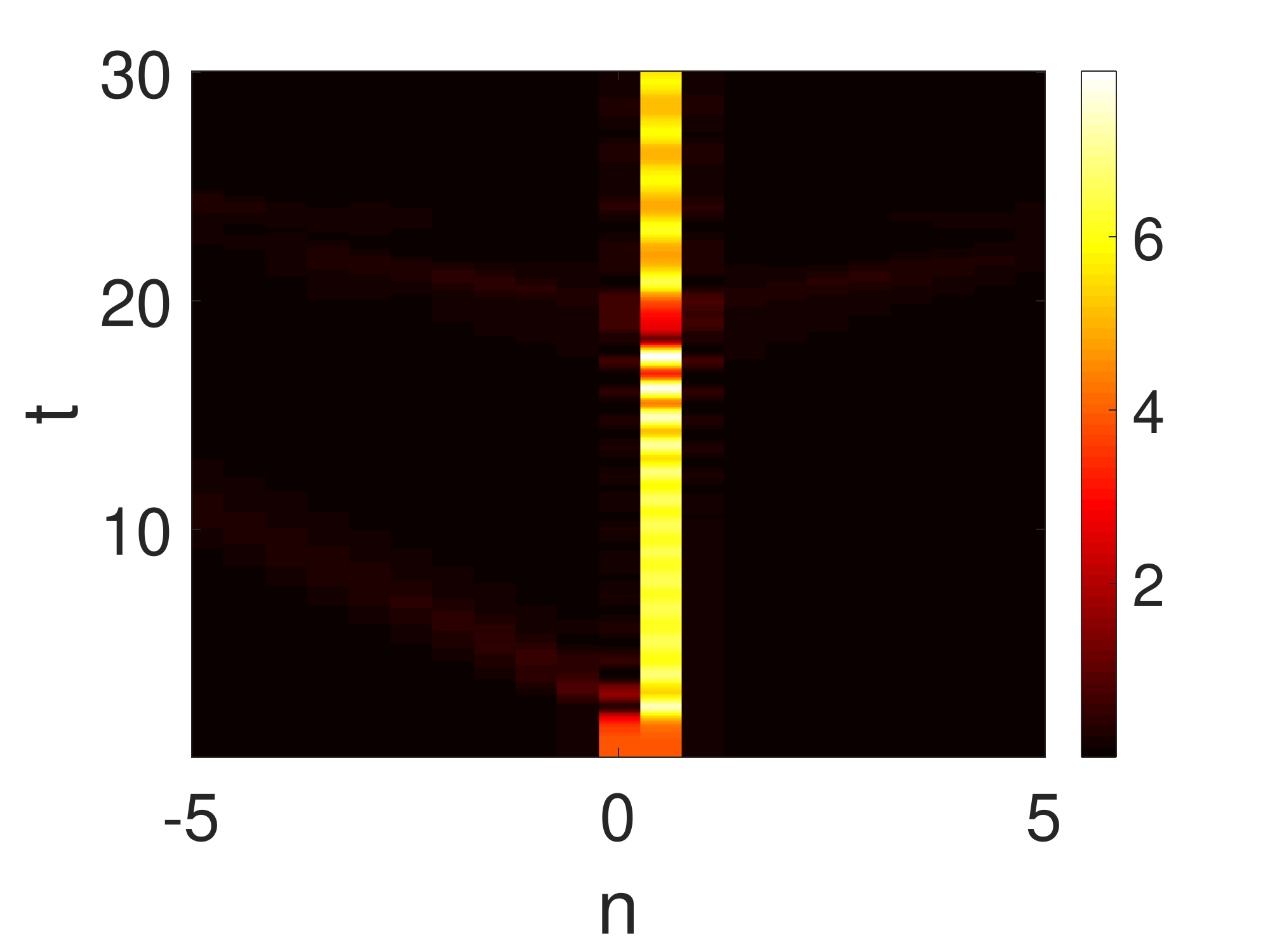}			\hspace{1cm}
\includegraphics[width = 0.35\textwidth]{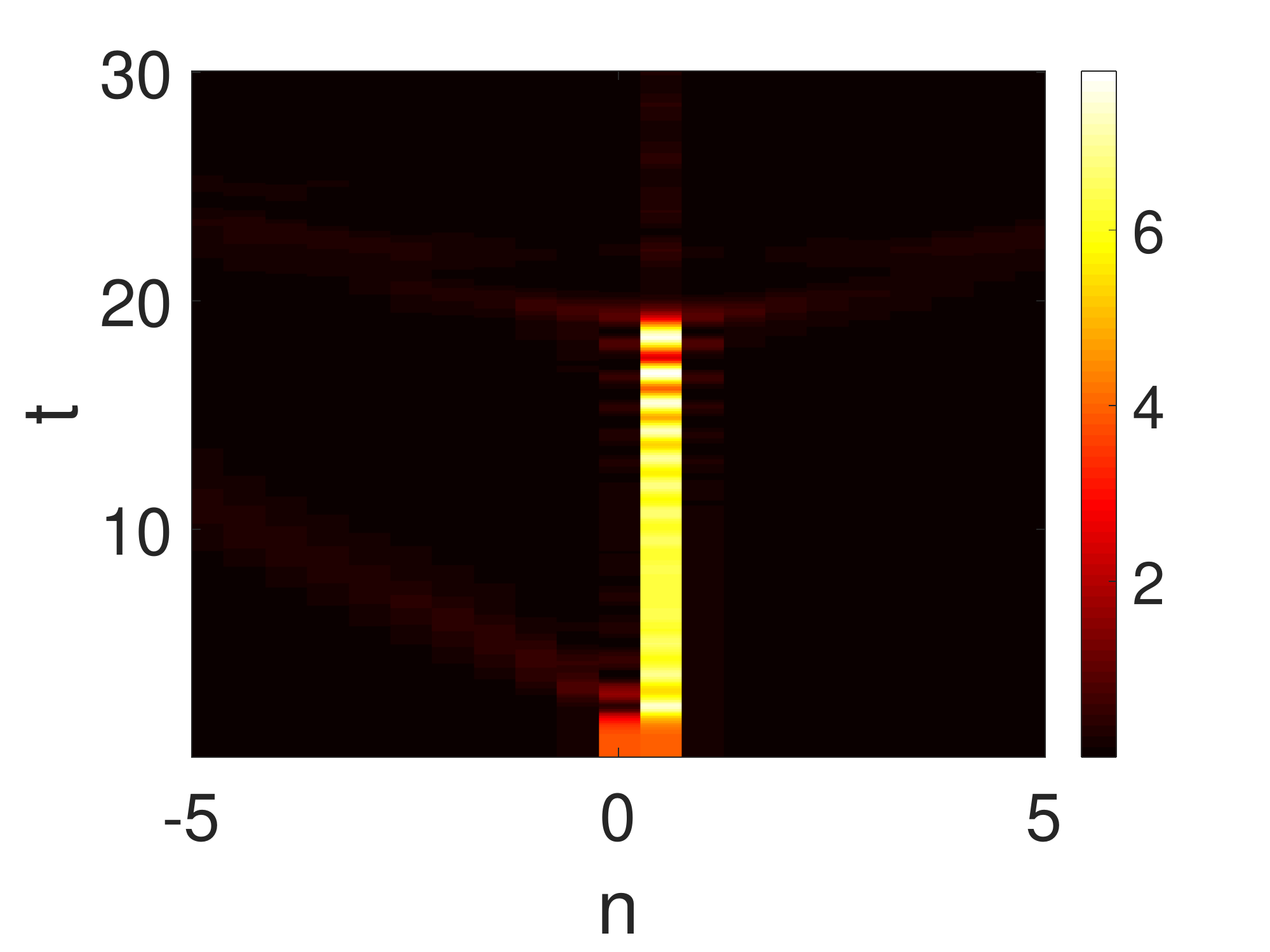}
\caption{Similar plots as Figure~\ref{fig.converge7}, they display the typical dynamics for an unstable antisymmetric intersite discrete soliton with the same parameter values (compare with Figure~\ref{fig.converge2}).}
\label{fig.converge8}
\end{figure*}

\begin{figure*}[tbhp!]
\centering
\includegraphics[width = 0.35\textwidth]{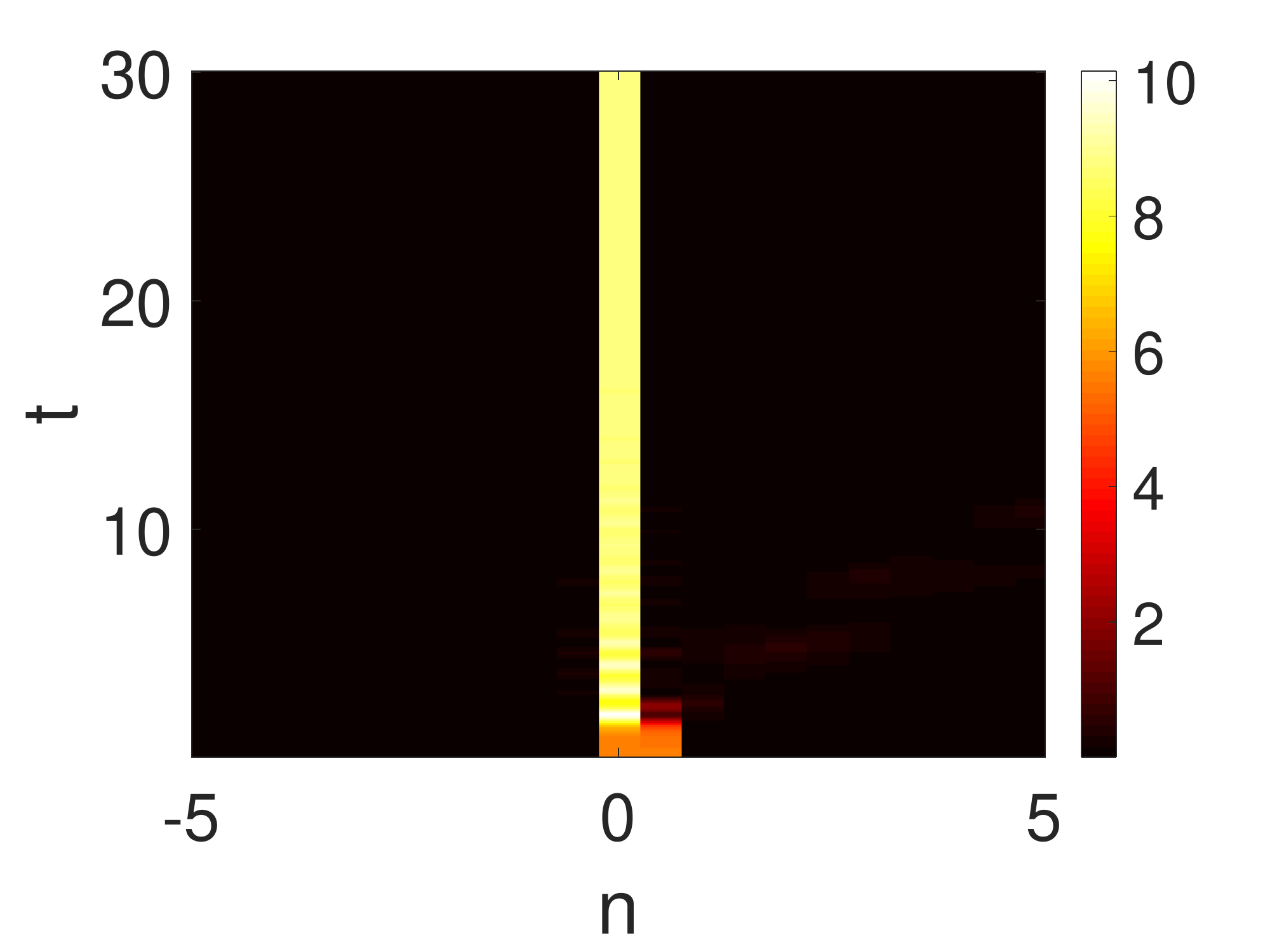}			\hspace{1cm}
\includegraphics[width = 0.35\textwidth]{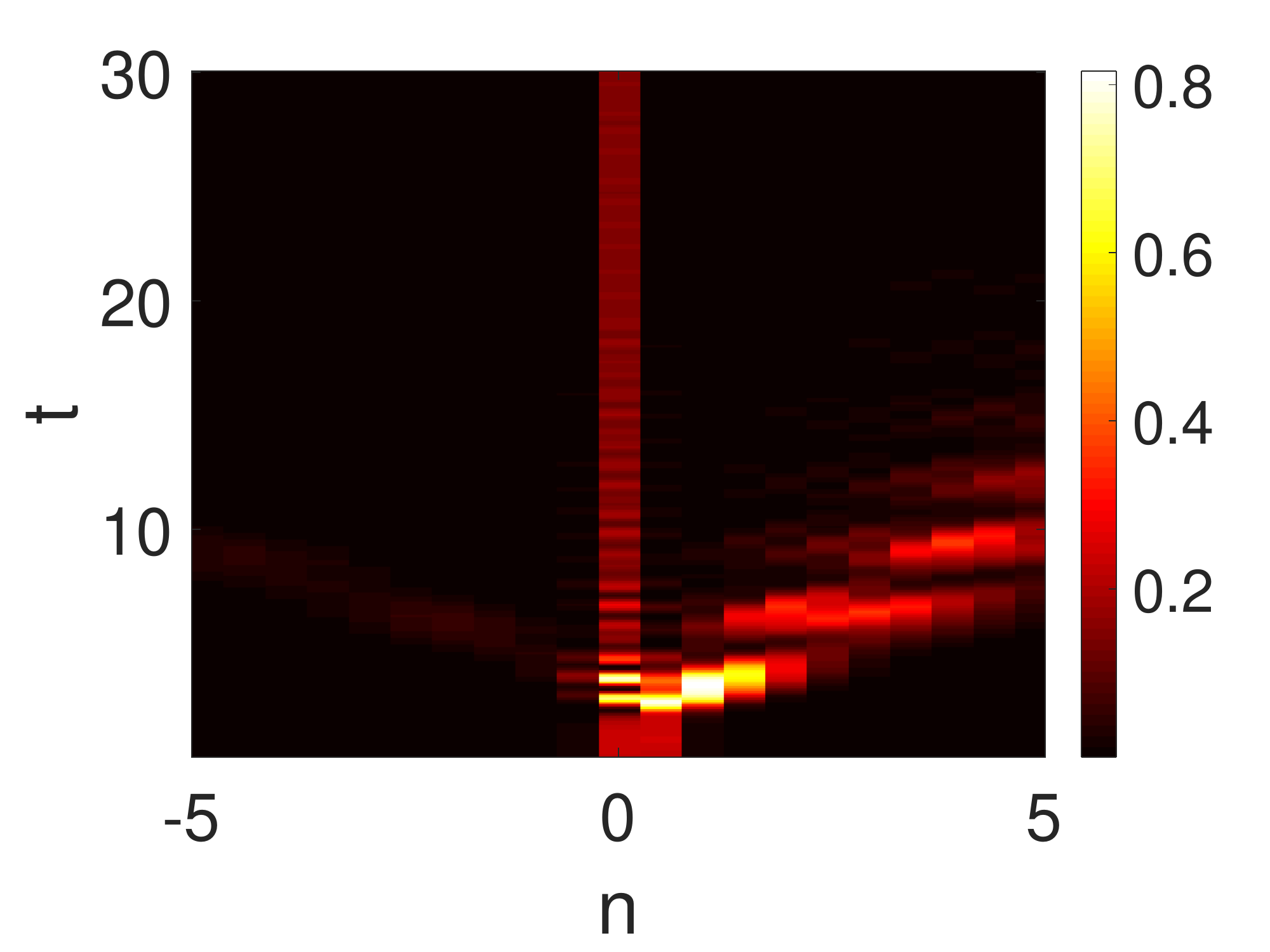}
\caption{Similar plots as Figure~\ref{fig.converge7}, they show the typical dynamics of unstable asymmetric intersite discrete solitons with $\omega = 5$, $\gamma = 0.5$, $\epsilon = 1$ (compare with Figure~\ref{fig.converge3}).} \label{fig.converge9}
\end{figure*}

\begin{figure*}[tbhp!]
\centering
\includegraphics[width = 0.35\textwidth]{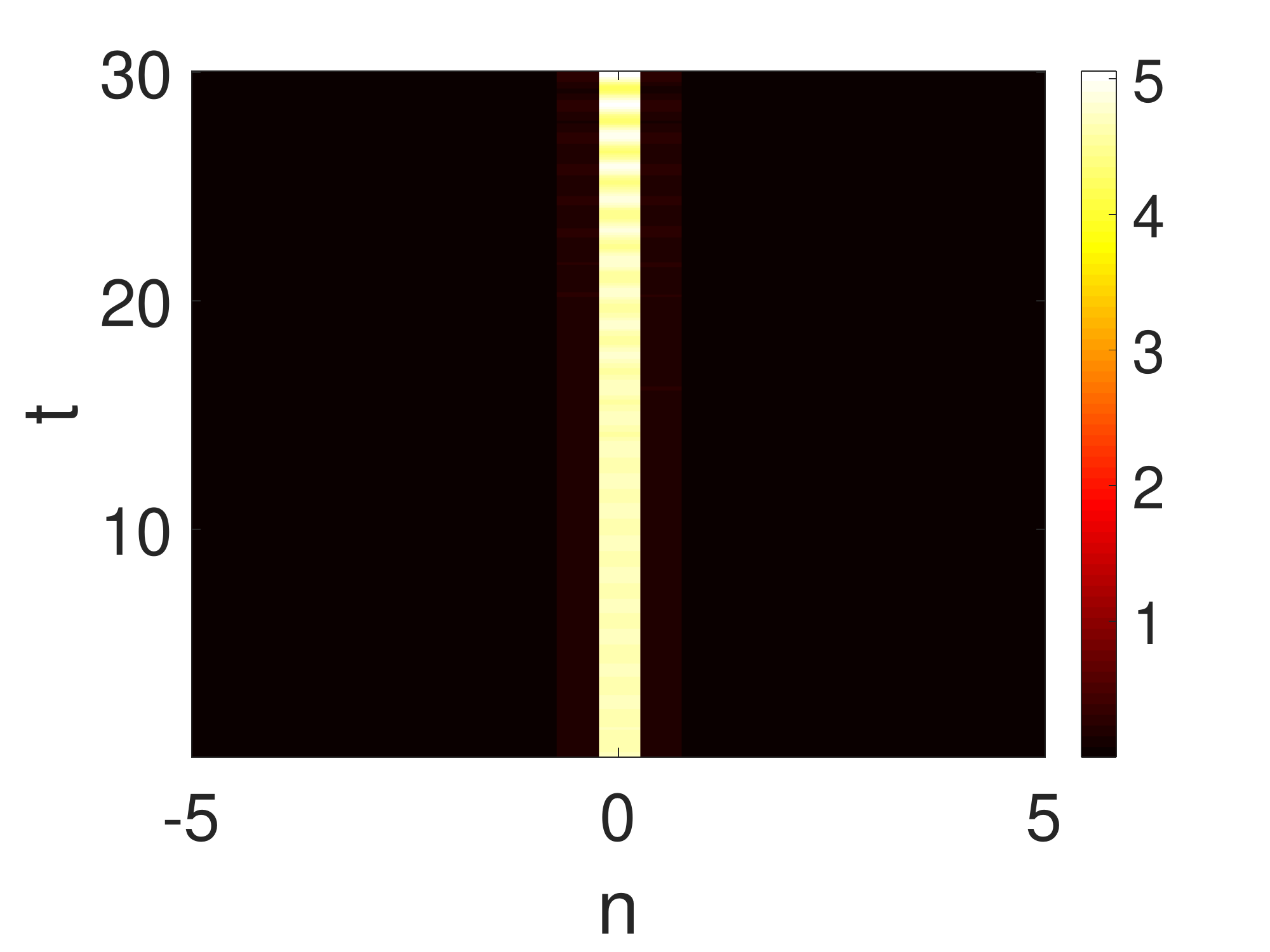}		 \hspace{1cm}
\includegraphics[width = 0.35\textwidth]{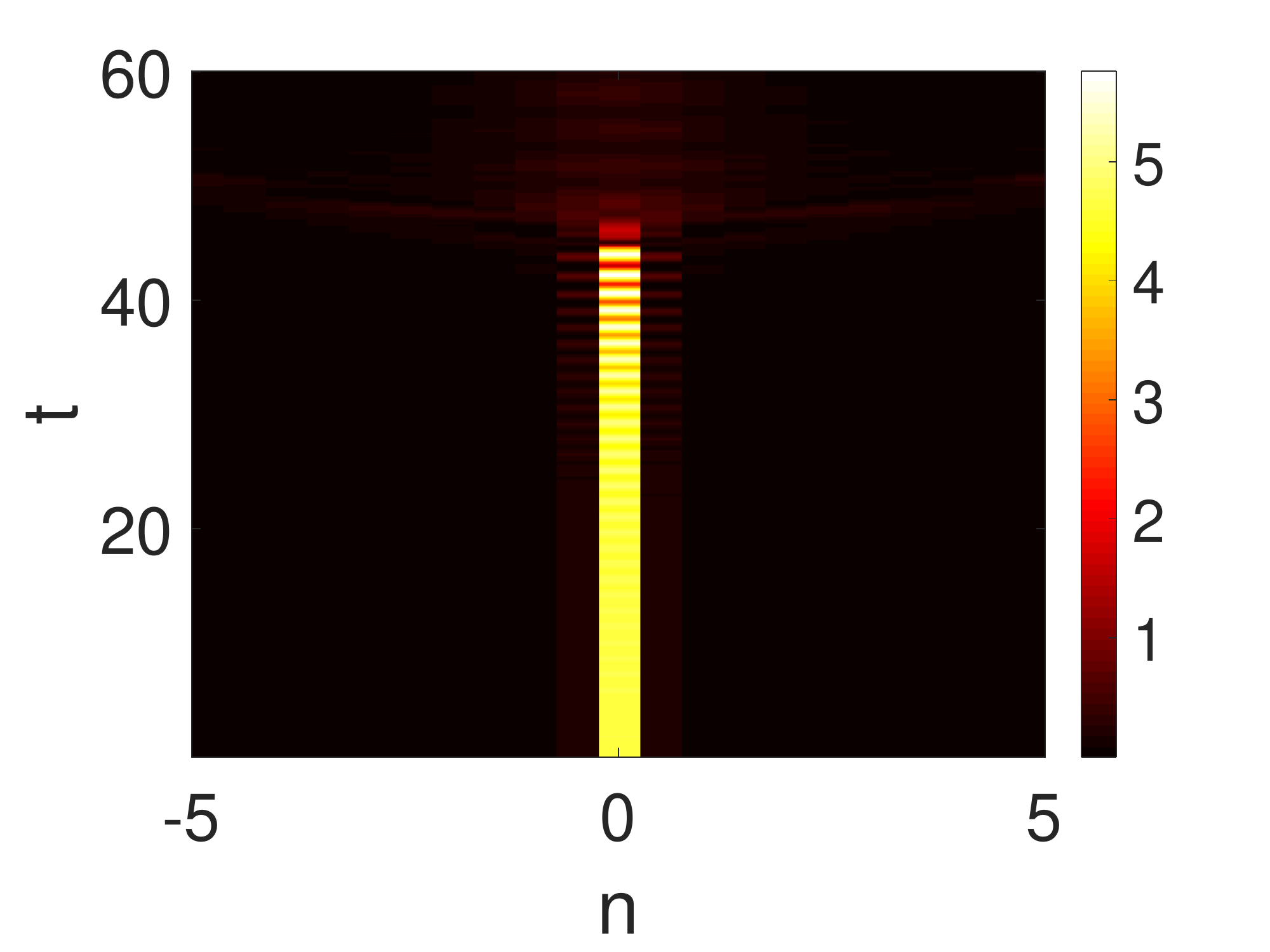}
\caption{Similar plots as Figure~\ref{fig.converge7}, they display the typical dynamics of unstable antisymmetric onsite discrete solitons with the same parameter values (compare with Figure~\ref{fig.converge5}).} \label{fig.converge10}
\end{figure*}

Finally, we present in Figures~\ref{fig.converge7}--\ref{fig.converge10} the time dynamics of the unstable solutions shown in Figures~\ref{fig.converge1}--\ref{fig.converge5}. We obtain one feature of typical dynamics in the form of discrete soliton destructions. One may attain oscillating solitons or asymmetric solutions between the arms. 

Similar to the families of discrete soliton in a \pts-symmetric chain of dimers with purely imaginary vertical coupling and real-valued velocity mismatch considered in~\cite{kirikchi2016bright}, most of the discrete solitons emanating from our model is also unstable, while the soliton families in a chain of dimers with $\cal{CP}$-symmetry considered in~\cite{kirikchi2018solitons} are stable. Stable discrete solitons in~\cite{kirikchi2016bright} occur when both the propagation constant and gain-loss parameter are small. On the other hand, the gain-loss coefficient does not influence the width of the snakes for the case \pts-symmetry chain of dimers with cubic-quintic nonlinearity~\cite{susanto2018snakes}.
% Figure 4

% Figure 5

% Figure 6
 
% Figure 7

% Figure 8

% Figure 9

% Figure 10

% end of figure
%%%%%%%%%%%%%%%%%%%%%%%%

% Section 5
\section{Conclusion}		\label{conclusion}

We have presented a model of double oligomers optical waveguide array using the discrete NLS equations with complex-valued coupling. The structure can be implemented in a discrete system with the \pts-symmetry characteristic. Both analytical and numerical results suggest the existence of fundamental bright discrete soliton solutions. We restricted our study to the two discrete modes of the solitons, the intersite and onsite modes. Furthermore, each mode possesses three distinct configurations between the arms of the dimers, depending on the real-valued amplitudes of the time-independent solution of the model in the anticontinuum limit. These are symmetric, asymmetric, and antisymmetric structures. 

We have also investigated the linear stability of the discrete soliton solutions by solving the corresponding linear eigenvalue problem. The continuous spectra lie on the imaginary axis and the parameter values determine the spectral boundaries. The corresponding discrete spectrum for the three structures of intersite discrete soliton admits one pair of eigenvalues bifurcating from the origin and two pairs of nonzero eigenvalues. On the other hand, for all three types of onsite discrete soliton, each structure possesses only one pair of nonzero discrete spectrum for small values of the horizontal linear coupling parameter. 

We observed the dynamics of the discrete spectra ranging from the anticontinuum to continuum limits, which correspond to an increasing value of the horizontal linear coupling parameter, for all the six types of discrete solitons. While all three types of intersite discrete solitons are always unstable, depending on the values of the propagation constant $\omega$ and the gain-loss parameter $\gamma$, onsite discrete solitons can be stable. A prevalent feature of the time dynamics for unstable discrete solitons is oscillation and annihilation as time progresses. We can extend to \pts-symmetric structure in a higher-dimension for future research.

% Acknowledgement
\begin{acknowledgements}
We are grateful to Professor Hadi Susanto, Department of Mathematics, Khalifa University, Abu Dhabi, The United Arab Emirates, for his assistance and valuable comments in improving this paper significantly.
\end{acknowledgements}

\section*{Conflict of interest}
The authors declare that they have no conflict of interest.

% References

\end{document}